\documentclass[fleqn,usenatbib]{config/mnras}

\usepackage[T1]{fontenc}
\usepackage{lastpage}

\DeclareRobustCommand{\VAN}[3]{#2}
\let\VANthebibliography\thebibliography
\def\thebibliography{\DeclareRobustCommand{\VAN}[3]{##3}\VANthebibliography}

\usepackage{amsmath}	% Advanced maths commands
\usepackage{graphicx}
\usepackage{booktabs}
\usepackage{multirow}	% Advanced maths commands 
\usepackage{paralist}
\usepackage{textgreek}
\usepackage{xargs}
\usepackage[dvipsnames,hyperref]{xcolor}
\usepackage{xspace}
\usepackage{txfonts}
\usepackage{hyperref}
\hypersetup{
    colorlinks=true,
    linkcolor=Cerulean,
    citecolor=NavyBlue,
    filecolor=magenta,
    urlcolor=Violet,
    pdftitle={26395 20415 ulema Francesco D'Eugenio WIDE JWST},
    pdfpagemode=FullScreen,
    }
\usepackage{subcaption}
\usepackage{adjustbox}  % For the orcid ids.

\graphicspath{{./}{figs/}}
\usepackage{etoolbox}
\makeatletter
\newcommand\sendemail[4]{%                %\newcommand\tpj@compose@mailto[3]{%
\edef\@tempa{mailto:#1?subject=#2&body=#3 }%
\edef\@tempb{\expandafter\html@spaces\@tempa\@empty}%
\href{\@tempb}{#4}}

\catcode\%=11
\def\html@spaces#1 #2{#1%20\ifx#2\@empty\else\expandafter\html@spaces\fi#2}
\catcode\%=14
\makeatother
\newcommand{\todo}[1]{\textcolor{magenta}{[#1]}}
\newcommand{\orcid}[2]{\href{http://orcid.org/#2}{#1}}
\newcommand{\orcidsymb}[2]{\href{http://orcid.org/#2}{#1\adjustbox{trim={-.15\width} {0\height} {-.15\width} {0\height},clip}{\includegraphics[height=10pt]{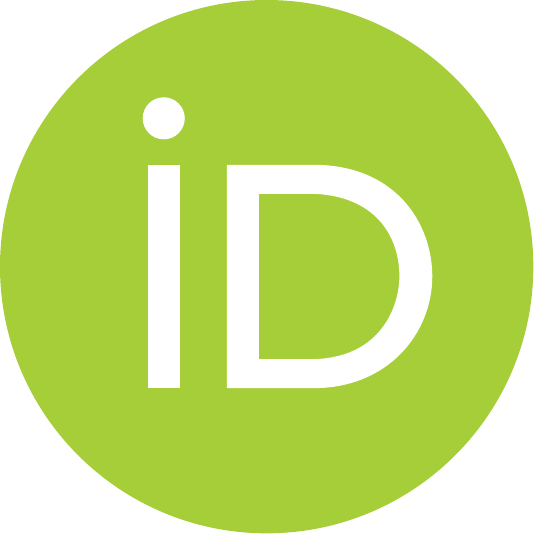}}}}

\newcommand{\citationneeded}{\textcolor{ForestGreen}{$^{\rm citation\;needed}$}}
\let\oldtextsigma\textsigma
\renewcommand{\textsigma}{\oldtextsigma\xspace}
\let\oldAA\AA
\renewcommand{\AA}{\text{\oldAA}\xspace}
\let\oldtextdegree\textdegree
\renewcommand{\textdegree}{\oldtextdegree\xspace}

\newcommand{\kms}{\ensuremath{\mathrm{km\,s^{-1}}}\xspace}
\newcommand{\Msun}{\ensuremath{{\rm M}_\odot}\xspace}
\newcommand{\Zsun}{\ensuremath{{\rm Z}_\odot}\xspace}
\newcommand{\yr}{\ensuremath{{\rm yr}}\xspace}
\newcommand{\Myr}{\ensuremath{{\rm Myr}}\xspace}
\newcommand{\Gyr}{\ensuremath{{\rm Gyr}}\xspace}
\newcommand{\peryr}{\ensuremath{{\rm yr^{-1}}}\xspace}
\newcommand{\Lsun}{\hbox{\,${\rm L}_\odot$}}
\newcommand{\mum}{\text{\textmu m}\xspace}
\newcommand{\kpc}{\text{kpc}\xspace}
\newcommand{\ZH}{\text{[Z/H]}\xspace}

\newcommandx{\lambdar}[2][1=R,2=]{\ensuremath{\lambda_{\rm {#1}}{#2}}\xspace}
\newcommand{\eps}{\ensuremath{\epsilon}\xspace}
\newcommand{\mstar}{\ensuremath{M_\star}\xspace}
\newcommand{\mdyn}{\ensuremath{M_\mathrm{dyn}}\xspace}
\newcommand{\re}{\ensuremath{R_\mathrm{e}}\xspace}
\newcommand{\vstar}{\ensuremath{v_\star}\xspace}
\newcommand{\vnai}{\ensuremath{v_{\NaI}}\xspace}
\newcommand{\sigmastar}{\ensuremath{\sigma_\star}\xspace}
\newcommand{\sigmaestar}{\ensuremath{\sigma_{\star,\mathrm{e}}}\xspace}
\newcommand{\vperc}[1]{\ensuremath{v_{#1}}\xspace}

\newcommand{\vesc}{\ensuremath{v_\mathrm{esc}}\xspace}
\newcommand{\nelec}{\ensuremath{n_\mathrm{e}}\xspace}
\newcommand{\Rout}{\ensuremath{R_\mathrm{out}}\xspace}
\newcommand{\vout}{\ensuremath{v_\mathrm{out}}\xspace}
\newcommandx{\Mout}[2][1=,2=]{\ensuremath{M_{\mathrm{out}{#2}}^{#1}}\xspace}
\newcommandx{\Mdotout}[2][1=,2=]{\ensuremath{\dot{M}_{\mathrm{out}{#2}}^{#1}}\xspace}

\newcommandx{\fluxdcgs}[1][1=-20]{$\times 10^{[#1]}$~erg~s$^{-1}$~cm$^{-2}$~\AA$^{-1}$\xspace}
\newcommandx{\fluxcgs}[2][1=-20,2=\ensuremath{\times}]{${#2}10^{#1}$~erg~s$^{-1}$~cm$^{-2}$\xspace}
\newcommandx{\powercgs}[1][1=44]{$\times 10^{#1}$~erg~s$^{-1}$\xspace}
\newcommand{\Av}{\ensuremath{A_V}\xspace}

\newcommand{\fHbetash}{$f(\mathrm{H}\text{\textbeta})_\mathrm{sh}$\xspace}
\newcommandx{\target}[1][1=]{\text{Ulema{#1}}\xspace}

\newcommand{\jwst}{\textit{JWST}\xspace}
\newcommand{\hst}{\textit{HST}\xspace}
\newcommand{\ppxf}{{\sc ppxf}\xspace}
\newcommand{\prospector}{{\sc prospector}\xspace}
\newcommand{\emcee}{{\sc emcee}\xspace}
\newcommand{\cloudy}{{\sc cloudy}\xspace}
\newcommand{\pyneb}{{\sc pyneb}\xspace}
\newcommandx{\mappings}[1][1=]{{\sc mappings{#1}}\xspace}
\newcommand{\shinbnzv}{{\sc shinbnzv}\xspace}

\newcommand{\Mdynvalue}{$\Mdyn = 2.0\pm0.5 \times 10^{11}$~\MSun}

\defcitealias{tacchella+2022a}{T22}
\defcitealias{nesvadba+2017}{N17}

\newcommand{\Lyalpha}{\text{Ly\textalpha}\xspace}
\newcommand{\Halpha}{\text{H\textalpha}\xspace}
\newcommand{\Hbeta}{\text{H\textbeta}\xspace}
\newcommand{\Hgamma}{\text{H\textgamma}\xspace}
\newcommand{\Hdelta}{\text{H\textdelta}\xspace}
\newcommand{\Paalpha}{\text{Pa\textalpha}\xspace}
\newcommand{\Pabeta}{\text{Pa\textbeta}\xspace}
\newcommand{\Hepsilon}{\text{H\textepsilon}\xspace}

\newcommandx{\permittedEL}[6][1=O,2=III,3=,4=,5=,6=]{\text{{#1}\,{\sc {#2}}{#3}{#4}{#5}{#6}}\xspace}
\newcommandx{\semiforbiddenEL}[6][1=O,2=III,3=,4=,5=,6=]{\text{{#1}\,{\sc{#2}}]{#3}{#4}{#5}{#6}}\xspace}
\newcommandx{\forbiddenEL}[6][1=O,2=III,3=,4=,5=,6=]{\text{[{#1}\,{\sc{#2}}]{#3}{#4}{#5}{#6}}\xspace}

\newcommand{\EW}[1]{\text{EW(#1)}\xspace}

\newcommand{\HI}{\permittedEL[H][i]}
\newcommand{\HII}{\permittedEL[H][ii]}

\newcommand{\NV}{\permittedEL[N][v]}
\newcommandx{\NVL}[1][1=1243]{\permittedEL[N][v][\textlambda][#1]}
\newcommandx{\NVall}{\permittedEL[N][v][\textlambda][\textlambda][1239,][1243]}

\newcommandx{\CIIL}[1][1=232x]{\semiforbiddenEL[C][ii][\textlambda][#1]}
\newcommandx{\CIIall}{\semiforbiddenEL[C][ii][\textlambda][\textlambda][2324--][2329]}

\newcommand{\NIV}{\semiforbiddenEL[N][iv]}
\newcommandx{\NIVL}[1][1=1486]{\semiforbiddenEL[N][iv][\textlambda][#1]}

\newcommand{\CIV}{\permittedEL[C][iv]}
\newcommandx{\CIVL}[1][1=1550]{\permittedEL[C][iv][\textlambda][#1]}
\newcommand{\CIVall}{\permittedEL[C][iv][\textlambda][\textlambda][1548,][1551]}

\newcommand{\HeII}{\permittedEL[He][ii]}
\newcommandx{\HeIIL}[1][1=1640]{\permittedEL[He][ii][\textlambda][#1]}

\newcommand{\semiOIII}{\semiforbiddenEL[O][iii]}
\newcommandx{\semiOIIIL}[1][1=1666]{\semiforbiddenEL[O][iii][\textlambda][#1]}
\newcommand{\semiOIIIall}{\semiforbiddenEL[O][iii][\textlambda][\textlambda][1661,][1666]}

\newcommand{\NIII}{\semiforbiddenEL[N][iii]}
\newcommandx{\NIIIL}[1][1=1750]{\semiforbiddenEL[N][iii][\textlambda][#1]}
\newcommand{\NIIIall}{\semiforbiddenEL[N][iii][\textlambda][\textlambda][1747--][1754]}

\newcommandx{\CIII}{\semiforbiddenEL[C][iii]}
\newcommandx{\CIIIL}[1][1=1909]{\semiforbiddenEL[C][iii][\textlambda][#1]}
\newcommand{\CIIIall}{\semiforbiddenEL[C][iii][\textlambda][\textlambda][1907,][1909]}

\newcommand{\NeIV}{\forbiddenEL[Ne][iv]}
\newcommandx{\NeIVL}[1][1=2424]{\forbiddenEL[Ne][iv][\textlambda][#1]}
\newcommand{\NeIVall}{\forbiddenEL[Ne][iv][\textlambda][\textlambda][2422,][2424]}

\newcommand{\MgII}{\permittedEL[Mg][ii]}
\newcommandx{\MgIIL}[1][1=2803]{\permittedEL[Mg][ii][\textlambda][#1]}
\newcommand{\MgIIall}{\permittedEL[Mg][ii][\textlambda][\textlambda][2796,][2803]}

\newcommand{\NeV}{\forbiddenEL[Ne][v]}
\newcommandx{\NeVL}[1][1=3426]{\forbiddenEL[Ne][v][\textlambda][#1]}
\newcommand{\NeVall}{\forbiddenEL[Ne][v][\textlambda][\textlambda][3346,][3426]}

\newcommand{\OII}{\forbiddenEL[O][ii]}
\newcommandx{\OIIL}[1][1=3727]{\forbiddenEL[O][ii][\textlambda][#1]}
\newcommand{\OIIall}{\forbiddenEL[O][ii][\textlambda][\textlambda][3726,][3729]}

\newcommand{\NeIII}{\forbiddenEL[Ne][iii]}
\newcommandx{\NeIIIL}[1][1=3869]{\forbiddenEL[Ne][iii][\textlambda][#1]}
\newcommand{\NeIIIall}{\forbiddenEL[Ne][iii][\textlambda][\textlambda][3869,][3967]}

\newcommand{\OIII}{\forbiddenEL[O][iii]}
\newcommandx{\OIIIL}[1][1=5007]{\forbiddenEL[O][iii][\textlambda][#1]}
\newcommand{\OIIIall}{\forbiddenEL[O][iii][\textlambda][\textlambda][4959,][5007]}

\newcommandx{\NIL}[1][1=5200]{\forbiddenEL[N][i][\textlambda][#1]}
\newcommand{\NIall}{\forbiddenEL[N][i][\textlambda][\textlambda][5198,][5200]}

\newcommand{\OI}{\forbiddenEL[O][i]}
\newcommandx{\OIL}[1][1=6300]{\forbiddenEL[O][i][\textlambda][#1]}
\newcommand{\OIall}{\forbiddenEL[O][i][\textlambda][\textlambda][6300,][6364]}

\newcommand{\HeI}{\permittedEL[He][i]}
\newcommandx{\HeIL}[1][1=5875]{\permittedEL[He][i][\textlambda][#1]}

\newcommand{\NII}{\forbiddenEL[N][ii]}
\newcommandx{\NIIL}[1][1=6584]{\forbiddenEL[N][ii][\textlambda][#1]}
\newcommand{\NIIall}{\forbiddenEL[N][ii][\textlambda][\textlambda][6549,][6584]}

\newcommand{\SII}{\forbiddenEL[S][ii]}
\newcommand{\SIIL}[1][1=6716]{\forbiddenEL[S][ii][\textlambda][#1]}
\newcommand{\SIIall}{\forbiddenEL[S][ii][\textlambda][\textlambda][6716,][6731]}

\newcommandx{\OIIAuL}[1][1=7325]{\forbiddenEL[O][ii][\textlambda][#1]}
\newcommand{\OIIAuall}{\forbiddenEL[O][ii][\textlambda][\textlambda][7319--][7331]}

\newcommandx{\CIIFIRL}{\forbiddenEL[C][ii][\textlambda][158\,\mum]}

\newcommand{\hda}{\ensuremath{\mathrm{H\text{\textdelta}_A}}\xspace}
\newcommand{\hga}{\ensuremath{\mathrm{H\text{\textgamma}_A}}\xspace}

\title[\jwst shock emission at z=4.6]{\jwst/NIRSpec WIDE survey: a z=4.6 low-mass star-forming galaxy hosting a jet-driven shock with low ionisation and solar metallicity}

\author[\sendemail{francesco.deugenio@gmail.com}{Questions about your WIDE paper}{Dear Francesco,\%0A\%0Ahow are you? I have a question about the paper , if I may.\%0AThe thing is, that ...\%0A\%0ARegards,\%0A}{F. D'Eugenio}~et al.]{\parbox{\textwidth}{
\orcidsymb{Francesco D'Eugenio}{0000-0003-2388-8172}$^{\hyperlink{aff1}{1},\hyperlink{aff2}{2}}$\thanks{E-mail: francesco.deugenio@gmail.com},
\orcidsymb{Roberto Maiolino}{0000-0002-4985-3819}$^{\hyperlink{aff1}{1},\hyperlink{aff2}{2},\hyperlink{aff3}{3}}$,
\orcidsymb{Vijay H. Mahatma}{0000-0001-5221-2636}$^{\hyperlink{aff2}{2}}$,
\orcidsymb{Giovanni Mazzolari}{}$^{\hyperlink{aff1}{1},\hyperlink{aff4}{4},\hyperlink{aff5}{5}}$,
\orcidsymb{Stefano Carniani}{0000-0002-6719-380X}$^{\hyperlink{aff6}{6}}$,
\orcidsymb{Anna de Graaff}{0000-0002-2380-9801}$^{\hyperlink{aff7}{7}}$,
\orcidsymb{Michael V.~Maseda}{0000-0003-0695-4414}$^{\hyperlink{aff8}{8}}$,
\orcidsymb{Eleonora Parlanti}{0000-0002-7392-7814}$^{\hyperlink{aff6}{6}}$,
\orcidsymb{Andrew J. Bunker}{0000-0002-8651-9879}$^{\hyperlink{aff9}{9}}$,
\orcidsymb{Xihan Ji}{0000-0002-1660-9502}$^{\hyperlink{aff1}{1},\hyperlink{aff2}{2}}$,
\orcidsymb{Gareth C.~Jones}{0000-0002-0267-9024}$^{\hyperlink{aff9}{9}}$,
\orcidsymb{Raffaella Morganti}{0000-0002-9482-6844}$^{\hyperlink{aff10}{10},\hyperlink{aff11}{11}}$,
\orcidsymb{Jan Scholtz}{0000}$^{\hyperlink{aff1}{1},\hyperlink{aff2}{2}}$,
\orcidsymb{Sandro Tacchella}{0000-0002-8224-4505}$^{\hyperlink{aff1}{1},\hyperlink{aff2}{2}}$,
\orcidsymb{Clive Tadhunter}{0000-0002-2951-3278}$^{\hyperlink{aff12}{12}}$,
\orcidsymb{Hannah \"Ubler}{
0000-0003-4891-0794}$^{\hyperlink{aff1}{1},\hyperlink{aff2}{2}}$ and 
\orcidsymb{Giacomo Venturi}{0000-0001-8349-3055}$^{\hyperlink{aff6}{6}}$
}\vspace{0.4cm}
\\
\parbox{\textwidth}{
\hypertarget{aff1}{$^{1}$}Kavli Institute for Cosmology, University of Cambridge, Madingley Road, Cambridge, CB3 0HA, United Kingdom\\
\hypertarget{aff2}{$^{2}$}Cavendish Laboratory - Astrophysics Group, University of Cambridge, 19 JJ Thomson Avenue, Cambridge, CB3 0HE, United Kingdom\\
\hypertarget{aff3}{$^{3}$}Department of Physics and Astronomy, University College London, Gower Street, London WC1E 6BT, UK\\
\hypertarget{aff4}{$^{4}$}Dipartimento di Fisica e Astronomia, Università di Bologna, Via Gobetti 93/2, I-40129 Bologna, Italy\\
\hypertarget{aff4}{$^{5}$}INAF – Osservatorio di Astrofisica e Scienza dello Spazio di Bologna, Via Gobetti 93/3, I-40129 Bologna, Italy\\
\hypertarget{aff6}{$^{6}$}Scuola Normale Superiore, Piazza dei Cavalieri 7, I-56126 Pisa, Italy\\
\hypertarget{aff7}{$^{7}$}Max-Planck-Institut f\"ur Astronomie, K\"onigstuhl 17, D-69117, Heidelberg, Germany\\
\hypertarget{aff8}{$^{8}$}Department of Astronomy, University of Wisconsin-Madison, 475 N. Charter St., Madison, WI 53706 USA\\
\hypertarget{aff9}{$^{9}$}Department of Physics, University of Oxford, Denys Wilkinson Building, Keble Road, Oxford OX1 3RH, UK\\
\hypertarget{aff10}{$^{10}$}ASTRON, the Netherlands Institute for Radio Astronomy, Oude Hoogeveensedijk 4, 7991 PD, Dwingeloo, The Netherlands\\
\hypertarget{aff11}{$^{11}$}Kapteyn Astronomical Institute, University of Groningen, Postbus 800, 9700 AV Groningen, The Netherlands\\
\hypertarget{aff12}{$^{12}$}Department of Physics and Astronomy, University of Sheffield, Sheffield, S7 3RH, UK\\
}
}

\date{Accepted XXX. Received YYY; in original form ZZZ}

\pubyear{2024}

\begin{document}
\label{firstpage}
\pagerange{\pageref{firstpage}--\pageref{lastpage}}
\maketitle

\begin{abstract}
We present NIRSpec/MSA observations from the \jwst large-area survey WIDE, targeting the rest-frame UV--optical spectrum of \target, a radio-AGN host at redshift $z=4.6348$. The low-resolution prism spectrum displays high equivalent width nebular emission, with remarkably high ratios of low-ionisation species of oxygen, nitrogen and sulphur, relative to hydrogen; auroral O$^+$ emission is clearly detected, possibly also C$^+$. From the high-resolution grating spectrum, we measure a gas velocity dispersion of $\sigma\sim400$~\kms, broad enough to rule out star-forming gas in equilibrium in the gravitational potential of the galaxy.
Emission-line ratio diagnostics suggest that the nebular emission is due to a shock which ran out of pre-shock gas.
To infer the physical properties of the system, we model simultaneously the galaxy spectral energy distribution (SED) and shock-driven line emission under a Bayesian framework.
We find a relatively low-mass, star-forming system ($\mstar = 1.4 \times 10^{10}~\Msun$, $\mathrm{SFR} = 70~\Msun~\peryr$), where shock-driven emission contributes 50~per cent to the total \Hbeta luminosity.
The nebular metallicity is near solar -- three times higher than that predicted by the mass-metallicity relation at $z=4.6$, possibly related to fast-paced chemical evolution near the galaxy nucleus.
We find no evidence for a recent decline in the SFR of the galaxy, meaning that, already at this early epoch, fast radio-mode AGN feedback was poorly coupled with the bulk of the star-forming gas; therefore, most of the feedback energy must end up in the galaxy halo, setting the stage for future quenching.
\end{abstract}


\section{Introduction}

There are many physical processes that can drastically reduce (or `quench') star formation in galaxies, making them (and keeping them) quiescent \citep[or `passive'; e.g.,][]{man+belli_quenching_2018}.
For high-mass galaxies, quenching and quiescence are likely caused by active galactic nuclei (AGN), through feedback from accreting supermassive black holes (SMBH). This hypothesis was first proposed based on simple energy arguments \citep{silk+rees1998,haehnelt+1998,binney2004}. In the meantime, supporting evidence has accumulated both from theory \citep{croton+2006,bower+2012,cresci+maiolino2018,harrison+2018,piotrowska+2022} and observations \citep{bluck+2022,belli+2023,deugenio+2023c,davies+2024}.

However, even if there is growing evidence for AGN-driven quenching, there is still the open problem of how exactly AGN quench star formation. Statistical studies of observations and simulations show that quiescence is most closely linked to SMBH mass \citep[a tracer of the time-integrated SMBH accretion;][]{bluck+2022,piotrowska+2022,brownson+2022}, rather than AGN luminosity or Eddington ratio \citep{stanley+2015,scholtz+2018,ward+2022}. Evidence of a systematic metallicity difference between star-forming and quiescent galaxies implies that the last period of star formation proceeds with little or no influx of pristine gas \citep{peng+2015,
trussler+2020}.
This has been interpreted as evidence for `preventive' feedback \citep{bower+2006,bower+2008}, which accumulates feedback energy not in the galaxy, but in its surrounding halo.
This form of feedback is not effective at stopping ongoing star formation, but -- by heating the galaxy halo -- prevents cold gas from accumulating and fuelling future star formation.

The alternative possibility of `ejective' AGN feedback, in contrast to preventive feedback, removes the fuel of star formation through fast, multi-phase outflows \citep[e.g.,][]{morganti+2005,hopkins+2008,nesvadba+2008,feruglio+2010,harrison+2012,maiolino+2012,cicone+2015,cresci+2015}.
While there is no consensus on whether or not these outflows quench star formation \citetext{see, e.g., \citealp{forsterschreiber+2019,carniani+2016,maiolino+2017,perna+2020,scholtz+2020,scholtz+2021,lamperti+2021}},
recent results from \jwst observations of quiescent galaxies at redshifts $z=3\text{--}5$ suggest rapid star formation and fast AGN feedback at early cosmic epochs \citep[i.e., within the first few hundreds Myr after the Big Bang; e.g.][]{carnall+2023b,glazebrook+2024,degraaff+2024,carnall+2024}, which may need some form of ejective feedback.
A few studies report observing ejective AGN feedback in action at z$>$2 in massive, quenching galaxies \citep{belli+2023,deugenio+2023c,davies+2024}.

The suite of models that describe ejective feedback are generally broken into two main categories, where the feedback loop is driven either by intense radiation from the central AGN, known as radiative-mode feedback, or by jets, known as radio-mode feedback.
Jet-driven outflows have been observed in all gas phases \citep[e.g.,][]{morganti+2005,holt+2006,holt+2008,holt+2011,santoro+2020,murthy+2022}.
However, radio-mode feedback is thought to be mainly responsible for preventive feedback, due to its inefficient coupling with the star-forming disc of the host galaxy \citep[e.g.,][]{croton+2006}.%tadhunter?
 Yet, more recently, both theoretical and observational works have revealed that low-power radio jets at low inclination angles relative to the galaxy disc can significantly affect the gas disc of star-forming galaxies \citep[e.g., ][]{mukherjee+2016,mukherjee+2018,venturi+2021}, possibly suppressing star formation via increased turbulence \citep[e.g.,][]{mandal+2021,roy+2024}, in addition to suppression via the combination of ejective and preventive feedback \citep{morganti+2021}.
Nevertheless, connecting AGN and quenching directly has proven challenging, due in part to timing arguments \citep[e.g.,][]{harrison+2018}, and in part to the co-fuelling of both SMBHs and star formation when cold gas is present \citep{vito+2014,ward+2022}.

Near-infrared spectroscopic studies of AGN-driven radio jets at redshifts $z=1\text{--}4$ show
that these systems are invariably found in gas-rich galaxies, with complex, multi-component emission, i.e. arising from both the host galaxy, the AGN, and shocked material \citep[][hereafter: \citetalias{nesvadba+2017}]{nesvadba+2017}.
These remarkable works pushed the capabilities of ground-based spectrographs to their limits, ultimately hitting two instrumental limitations; a limited wavelength coverage, and the inability to probe the strongest emission lines beyond redshift $z=2\text{--}3$.
At the same time, the lack of suitable models to capture the multi-component nature of these complex sources hampered a clear view of their physical properties.
In this work, we show that all these limitations can be addressed to provide new insights into the nature of AGN feedback at high redshift.

With the launch of \jwst, we can finally study distant radio AGN by exploring their full range of rest-frame optical emission lines \citep[e.g.,][]{saxena+2024,roy+2024}.
The NIRSpec GTO programme WIDE aims to observe a large sample of bright galaxies at redshifts $z=1\text{--}7$; the survey goals include both statistical studies of the bright-end of the luminosity function, as well as the study of rare or short-lived phenomena \citep{maseda+2024}.
At the same time, modelling of the spectral energy distribution (SED) of galaxies has seen tremendous progress, with the development of many flexible tools \citetext{e.g., \textsc{beagle}, \citealp{Chevallard16}; \textsc{bagpipes}, \citealp{carnall+2019}; and \textsc{cigale}, \citealp{bouquien+2019}; to name a few}.
Modelling the emission of shocked gas is only a matter of correctly implementing existing shock models \citep[e.g.,][]{allen+2008,alarie+morisset2019} into these existing SED-fitting frameworks.

In this work, we study the emission-line and stellar-population properties of a radio-loud AGN and its host at $z=4.6$. Thanks to WIDE's combination of low- and high-spectral-resolution \jwst/NIRSpec spectroscopy (\S~\ref{s.obs}), we are able to study the ionisation, metallicity and kinematic properties of the gas in detail. We do so using a novel implementation that marries existing galaxy spectro-photometric modelling
\citep[\prospector;][]{johnson+2021} to shock-driven nebular emission \citetext{\mappings[~v], \citealp{sutherland+dopita2017}; via the 
\href{https://sites.google.com/site/mexicanmillionmodels/}{Mexican Million Models Database, 3MdB,} \citealp{alarie+morisset2019}}.
With this new approach, we simultaneously estimate the physical properties of the shock and those of the galaxy host
(\S~\ref{s.host}), finding a puzzling combination of a strong shock but a relatively low-mass host.
We discuss the implications of our findings in the context of jet-driven quenching (\S~\ref{s.disc}), and close with a succinct summary in \S~\ref{s.conc}.

Throughout this work, we assume the \citet{planck+2020} cosmology, which gives a physical scale of 6.66~kpc~arcsec$^{-1}$ at the redshift of the source, $z=4.6348$. All masses are total stellar mass formed. We use everywhere a \citet{chabrier2003} initial mass function, integrated between 0.1 and 120~\Msun; all wavelengths are vacuum wavelengths, but optical emission lines are named after their air wavelengths. All equivalent widths are in the rest frame.
We adopt a mass-weighted solar metallicity $\Zsun = 0.0142$, unless otherwise specified.

\begin{figure*}
   \includegraphics[width=\textwidth]{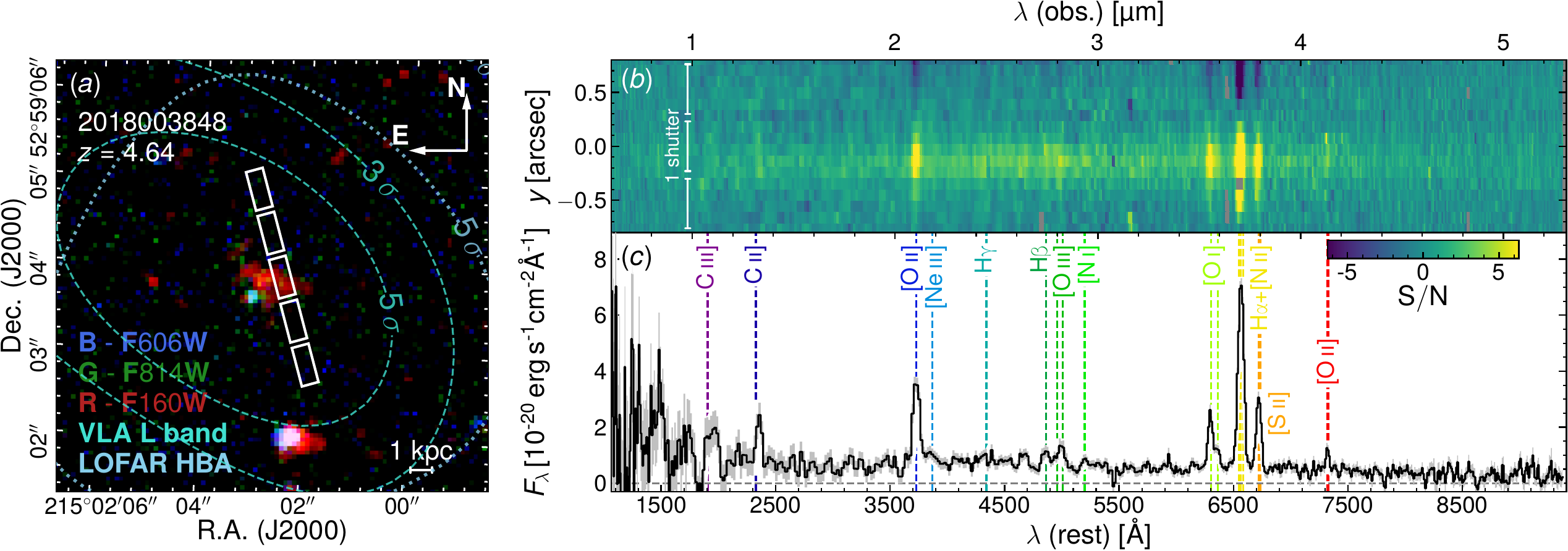}
   {\phantomsubcaption\label{f.img.a}
    \phantomsubcaption\label{f.img.b}
    \phantomsubcaption\label{f.img.c}}
   \caption{\hst/ACS and WFC3 false-colour image (panel~\subref{f.img.a}), showing rest-frame UV photometry of \target. The NIRSpec/MSA shutters are overlaid in white, the cyan dashed lines are the 3- and 5-\textsigma contours from VLA 1.4~GHz (rms=30~\textmu Jy; the beam size is 3.8~arcsec), and the dotted line
   is the 5-\textsigma contour from LOFAR 144~GHz (rms=60~\textmu Jy; beam size 6~arcsec).
   The source is not centred on the shutters; also note the bright UV region south-east of the galaxy.
   It is unclear whether this emission is associated with the target or if it is from an interloper, but its absence in F435W is consistent with the \Lyalpha drop at the redshift of \target. Panel~\subref{f.img.b} shows the 2-d signal-to-noise map of the NIRSpec/MSA prism spectrum, highlighting the extended nature of this galaxy. The 1-d spectrum (panel~\subref{f.img.c}) shows several low-ionisation and neutral lines (\OIIall, \OIall, \SIIall, \OIIAuall, possibly \CIIall) and weak high-ionisation lines (\OIIIall), revealing the nebular emission is powered by shocks (cf.~Fig.~\ref{f.bpt})}\label{f.img}
\end{figure*}

\begin{figure}
   \includegraphics[width=\columnwidth]{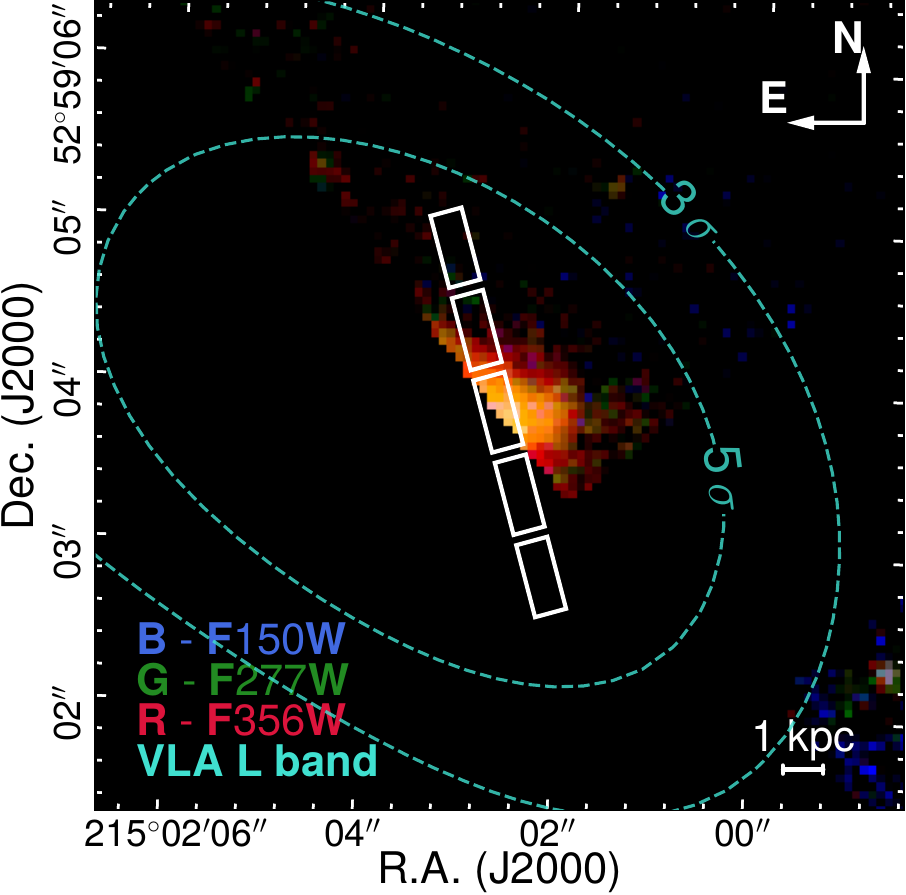}
   \caption{\jwst/NIRCam false-colour image of the target, from CEERS \citep{bagley+2023}. In the rest-frame optical, the target tentatively displays a smoother structure than in the rest-frame UV (cf.~Fig.~\ref{f.img.a}). Because the target centre lies outside of the NIRCam footprint, we do not use NIRCam photometry in the analysis. The VLA L-band contours are the same as in Fig.~\ref{f.img.a}.}\label{f.nircam}
\end{figure}

\section{Observations and data analysis}\label{s.obs}

Our target WIDE-AEGIS-2018003848 \citetext{R.A.: 215.034109, Dec: +52.984404; 3D-HST v4.1 ID: 26395, \citealp{brammer+2012,skelton+2014}; CANDELS ID: 20415, \citealp{grogin+2011,koekemoer+2011}} is drawn from the `AEGIS' observations of the WIDE survey \citep{maseda+2024}, which leverages existing
panchromatic observations of the `Extended Groth Strip' field \citep[EGS;][]{rhodes+2000,davis+2007},
obtained as part of the All-Wavelength Extended Groth Strip International Survey \citep[AEGIS;][]{davis+2007}.
The source was originally detected by \textit{Spitzer} as EGSIRAC~J142008.19+525903.7 \citep{barmby+2008}, and later associated with 1.4-GHz emission using the VLA \citep{willner+2012}.
It was observed by WIDE as a priority-1 target, based on being a radio AGN \citep[alongside other $\sim20$ radio AGN;][]{maseda+2024}, and we characterise it here as an `ultra low-ionisation emission-line AGN' (hereafter: \target).

\subsection{WIDE spectroscopy data}

WIDE uses the micro-shutter assembly (MSA) on \jwst/NIRSpec \citep{ferruit+2022,jakobsen+2022}, and observes each galaxy using a `slitlet' consisting of 3 shutters.
The spectrograph was configured with three disperser--filter combinations, the low-resolution PRISM/CLEAR (hereafter; prism), and the high-resolution configurations G235H/F170LP and G395H/F290LP.
The prism spectrum covers 0.7--5.3~\mum with spectral resolution $R=30\text{--}300$, while the two gratings cover together 1.66--5.3~\mum with $R=2,700$ \citep{jakobsen+2022}.
All spectra were observed using nodding; the prism observations use three nods and a total integration time of 41~min, whereas the grating spectra used only two nods and integration times of 27 and 30~min each for G235H and G395H, respectively.
We used the NRSIRS2 readout mode \citep{rauscher+2012,rauscher+2017}.
Our target belongs to the WIDE field number 18 (Programme ID~1213), and was observed on 23\textsuperscript{rd} March 2023. The observations were inspected for signatures of short circuits \citep{rawle+2022} and none were found.

The data reduction follows the procedures established in JADES \citetext{\jwst Advanced Deep Extragalactic Survey; \citealp{eisenstein_overview_2023}; \citealp{curtis-lake+2023,cameron+2023a,carniani+2023}, and \citealp{bunker+2023b}}.
We used a combination of the STScI and ESA NIRSpec pipelines \citep{alvesdeoliveira+2018,ferruit+2022}, as described in \citet{carniani+2023}.
In addition, the WIDE data reduction adds some survey-specific optimisations, described in \citet{maseda+2024}.
Crucially, the slit-loss corrections assume a point-source morphology \citep{bunker+2023b,curti+2023b}, therefore we rely on photometry to obtain unbiased colour estimates.

We note that in the G235H grating we have only one line detection of \OIIall, and it has low signal-to-noise ratio (SNR). Therefore, in the remainder of this article, we use only the G395H grating (hereafter, the grating).
We do not apply a Galactic foreground extinction correction, but note that at the location of this target, the Galaxy has $A_V = 0.03$~mag, implying a correction of 2--1 per~cent between 0.6--1~\mum, and even smaller at longer wavelengths.

The galaxy morphology and the prism spectrum are displayed in Fig.~\ref{f.img}. We obtained the reduced photometric images from the version~1 of the \href{https://dawn-cph.github.io/dja/index.html}{DAWN JWST Archive} \citep[DJA;][]{valentino+2023}.
Panel~\subref{f.img.a} shows rest-frame UV photometry from \hst/ACS and WFC3/IR, with overlaid the position of the NIRSpec shutters.
\target presents a complex morphology, consisting of several clumps. The MSA shutters do not cover the brightest clump, which may indicate the centre of the galaxy. However, the precise centre of the source is hard to determine in the rest-UV photometry. One of these clumps shows a distinctively blue colour.
\jwst/NIRCam imaging is available through the  Cosmic Evolution Early Release Science \citep[CEERS;][]{bagley+2023}, but only for part of the galaxy.
From these images (Fig.~\ref{f.nircam}), the galaxy seems significantly smoother and more extended at rest-frame optical wavelengths than in the rest-UV image from \hst.
All together, these properties suggest that the UV image is shaped primarily by the distribution of dust and/or by clumpy star formation.
Note that the CEERS observations do not fully cover the galaxy; the target centre (as estimated from \hst) seems to lie outside of the NIRCam footprint, which makes measuring accurate total fluxes even more challenging. For this reason, in the following we do not use NIRCam photometry, and supplement our spectroscopy observations with legacy data from \citet[][see our Table~\ref{t.phot}]{stefanon+2017}.

The 1-d spectral extraction (Fig.~\ref{f.img.c}) shows bright emission by low-ionisation species like \OIIall, \OIall, \SIIall, and \OIIAuall. To our knowledge, this is the most distant detection of the \OI and auroral \OII lines to date.
All these lines are bright compared to \OIIIall, which is a combination usually seen in high-metallicity, low-ionisation galaxies and in shock-dominated systems.
Another line group typically seen in shocks is the auroral \OIIAuall, which is clearly detected; these lines are coupled to tentative detections of \CIIall and \NIall.
The spectrum is clearly extended in the slit, and displays a clear gradient in both emission-line ratios and equivalent width (EW; e.g., at $y=-0.5$~arcsec \OIIall is fainter relative to the \Halpha+\NIIall complex, while EW(\Halpha+\NIIall) clearly increases from $y=0$ to $-0.5$~arcsec, due to the continuum becoming fainter).
These strong emission-line gradients mean that applying a photometry-based aperture correction introduces a bias against the average properties of the target.

\subsection{Ancillary data: photometry and radio observations}

We use photometry spanning from optical to near-infrared, obtained from the publicly available catalogue of \citet{stefanon+2017}. Our motivation for using photometry is to infer accurate aperture corrections by combining the total fluxes from photometry with the detailed physical information provided by the spectra.
This comparison is performed within the Bayesian inference framework of \prospector (\S~\ref{s.host.ss.prosp} and Appendix~\ref{a.sfonly}).

Because we do not model the resonant \Lyalpha line, we exclude the \hst/ACS F606W filter. Additionally,
we do not use the $J$-band upper limit from CFHT/WIRCAM (we have more informative detections from
\hst/WFC3 IR in the same wavelength region). Our photometry
selection is summarised in Table~\ref{t.phot}, and spans observed wavelengths between 0.8~\mum (with \hst/ACS) and 8~\mum (with \textit{Spitzer}/IRAC; aside from VLA and LOFAR, the longest-wavelength detection is at 4.5~\mum).

\begin{table}
    \centering
    \setlength{\tabcolsep}{4pt}
    Summary of literature photometry for \target.
    \begin{tabular}{llcc}
  \toprule
    Instrument & Filter & Flux         & Source \\
               &        & [\textmu Jy] &  \\
  \midrule
    \textit{{HST}}/ACS      & f606w$^\dagger$ & $0.05 \pm 0.01$ & \citep{koekemoer+2011} \\
    \textit{{HST}}/ACS      & f814w           & $0.15 \pm 0.02$ & " \\
    \textit{{HST}}/WFC3 IR  & f125w           & $0.18 \pm 0.02$ & " \\
    \textit{{HST}}/WFC3 IR  & f140w           & $0.31 \pm 0.05$ & \citep{skelton+2014} \\
    \textit{{HST}}/WFC3 IR  & f160w           & $0.25 \pm 0.03$ & \citep{koekemoer+2011} \\
    CFHT/WIRCAM             & J$^\dagger$     & $        <0.38$ & \citep{bielby+2012} \\
    CFHT/WIRCAM             & H               & $0.40 \pm 0.11$ & " \\
    CFHT/WIRCAM             & Ks              & $        <0.43$ & " \\
    \textit{{Spitzer}}/IRAC & 3.6 \mum        & $2.12 \pm 0.11$ & \citep{ashby+2015} \\
    \textit{{Spitzer}}/IRAC & 4.5 \mum        & $1.27 \pm 0.11$ & " \\
    \textit{{Spitzer}}/IRAC & 5.8 \mum        & $        <3.11$ & \citep{bielby+2012} \\
    \textit{{Spitzer}}/IRAC & 8.0 \mum        & $        <2.64$ & " \\
  \bottomrule
    \end{tabular}
    \caption{All measurements are from the publicly available table of \citet{stefanon+2017}; the Source column indicates the original source of the data. We do not use existing \jwst photometry because \target lies partially or completely outside of the NIRCam detector.
    $^\dagger$ Unused bands, see main text.
    }\label{t.phot}
\end{table}

This source is also detected at radio wavelengths by the VLA at 1.4~GHz, with a flux density $S_{1.4\,\mathrm{GHz}} = 0.056\pm0.011$~mJy \citep{ivison+2007,willner+2012}.
The source is also detected in the LOFAR Two-Metre Sky Survey \citep[LoTSS DR2;][]{shimwell+2022} at 144\,MHz at 6-arcsec resolution, with a flux density of $S_{144\,\mathrm{MHz}} = 0.37\pm0.037$~mJy\footnote{We have assumed a flux density error of 10~percent in line with typical errors in the survey release \citep{shimwell+2019}.}.
Based on the measurements at 144\,MHz and 1.4\,GHz, the radio spectral index (slope of the SED between 144 and 1.4 GHz, assuming a power-law energy distribution) is $-0.83\pm0.09$ (using the convention $S\propto\nu^{\alpha}$ where $\alpha$ is the spectral index).

From this value of $\alpha$, we infer $k$-corrected radio luminosity densities of
$P_{1.4\,\mathrm{GHz}} = 9.1\pm1.8 \times 10^{24}$~W~Hz$^{-1}$ and
$P_{144\,\mathrm{MHz}} = 6.0\pm0.6 \times 10^{25}$~W~Hz$^{-1}$ \citep[following the approach of ][]{pracy+2016}.
These values imply that \target is a low-luminosity radio galaxy. Nevertheless, the 1.4-GHz radio luminosity is two orders of magnitude above the star-formation dominated threshold 
\citep{pracy+2016}, implying that this source is dominated by AGN emission.
This is confirmed by the fact that the radio-inferred SFR clearly exceeds the value derived from \Halpha. Indeed, if the radio emission was dominated by star formation, then from $P_{1.4\,\mathrm{GHz}}$ we would estimate a SFR$_{1.4\,\mathrm{GHz}}$ of 300~\Msun~\peryr \citep{davies+2017}, while from $P_{144\,\mathrm{MHz}}$ we would infer $\mathrm{SFR}_{144\,\mathrm{MHz}}=300\text{--}1,000~\Msun~\peryr$ \citep[][or even higher, $\mathrm{SFR}_{144\,\mathrm{MHz}} \approx 2,000~\Msun~\peryr$ using either eq.~1 or 2 from \citealp{smith+2021}]{heesen+2022}. Both values are in excess of our aperture-corrected estimate of $\mathrm{SFR}_{10}\approx70~\Msun~\peryr$ (Table~\ref{t.prosp}), validating our assessment that the radio emission has an AGN origin.

The spectral index $\alpha=-0.83$ is typically characterised as `steep spectrum' at these frequencies for sources with an active jet \citep[e.g.,][]{zajacek+2019}: the high-energy components of radio galaxies (i.e. the jets and hotspots) are thought to have flat spectra at low radio frequencies \citep[$\alpha_\text{low}\approx-0.5$; associated with particle acceleration from strong shocks, e.g.,][]{meisenheimer+1989}. However, statistical studies of radio galaxies as a population have median spectral indices $\alpha_{150}^{1400}$ between -0.85 and -0.75 \citep[e.g.,][]{mahony+2016,kutkin+2023}, placing \target close to the population median. Given the uncertainties in the redshift evolution of $\alpha$ measured in radio surveys (due, e.g., to surface-brightness dimming at high redshift, redshift-dependent inverse-Compton losses, and intrinsic evolution), we explore the possibility that for \target we are seeing the old, leftover plasma from previous jet activity, or that the jet is recurrent \citep[consistent with studies on the radio spectra of low ionization AGN with star-forming activity, e.g.][]{zajacek+2019}. At the redshift of the source ($z=4.6$), we expect strong radiative losses at radio frequencies due to inverse-Compton scattering (low energy electrons up-scattering CMB photons), meaning a natural depletion of low energy electrons so that we ought to be sensitive only to the high energy, flat-spectrum components related to a jet. That we still observe a relatively steep radio spectrum (and not $\alpha\approx-0.5$) might suggest that the radio jet is no longer active, or is restarting, and we are seeing the leftover plasma. This would further support the conclusion drawn by the high flux ratio between low-ionisation and high-ionisation species, which suggests an old-shock origin (Section~\ref{s.sh} and Fig.~\ref{f.bpt}). However, given that we do not have robust measurements to determine an accurate radio spectral index at multiple radio frequencies, our interpretation is only tentative. Resolved observations at higher angular resolution are needed to determine whether an active jet exists in \target.

The target is undetected in X-rays, despite being covered by the deepest \textit{Chandra} X-ray observations available in the field \citep[AEGIS-XD, 800~ks;][]{nandra+2015}, implying an upper limit on the X-ray luminosity $L_\mathrm{X}\lesssim 5$~\fluxcgs[43].

\subsection{Emission-line fitting -- prism spectrum}\label{s.obs.ss.lowres}

We fit the emission lines using the \textchi$^2$-minimisation algorithm \ppxf \citep{cappellari2023}, which models simultaneously the emission lines and the stellar continuum. The emission lines are modelled as Gaussians, and the stellar continuum is a non-negative superposition of simple stellar population (SSP) spectra spanning a grid of ages and metallicities.
The SSP spectra employ MIST isochrones \citep{choi+2016} and C3K model atmospheres \citep{conroy+2019}.
Both the emission-line and SSP spectra have been pre-convolved to match the instrumental resolution of the prism spectrum \citep{jakobsen+2022}.
Because the size of our target is larger than the width of the MSA shutters, we use the nominal spectral resolution of each disperser \citetext{\citealp{jakobsen+2022}, i.e., we do not calculate an \textit{ad-hoc} line spread function, which would be appropriate for sources that are smaller than the shutters, \citealp{degraaff+2023}}.
The full list of emission lines and doublets we measure is presented in Table~\ref{t.emlines}.
Among these lines, we highlight the presence of \CIIall and \OIIAuall -- two auroral line groups characteristically associated with strong shocks \citep[e.g.,][their figs.~16 and 18]{allen+2008}.

\begin{table}
    \centering
    \setlength{\tabcolsep}{4pt}
    Redshifts and nebular emission line fluxes from the WIDE NIRSpec/MSA data of \target.
    \begin{tabular}{lllcc}
  \toprule
   \multirow{19}{*}{\rotatebox[origin=c]{90}{PRISM}}
   \multirow{19}{*}{\rotatebox[origin=c]{90}{\ppxf \S~\ref{s.obs.ss.lowres}}}
    & & Redshift                   &    ---       & $4.644\pm0.001$   \\
   \cline{3-5}
    & & \multirow{2}{*}{Line(s)}   & Wavelength   & Flux \\
    & &                            & [\AA vacuum] & [\fluxcgs[-18][]] \\
   \cline{3-5}
   & & \CIIIall$^\dagger$         & 1907.71 & $<3.3$ \\
   & & \CIIall$^\dagger$          & 2325.40 & $5.4\pm0.9$ \\
   & & \MgIIall$^\dagger$         & 2799.94 & $<2.4$ \\
   & & \OIIall$^\dagger$            & 3728.49 & $11.0\pm0.5$ \\
   & & \NeIIIall$^\ddag$          & 3870.86 & \multirow{2}{*}{$<1.5$} \ \\
   & & \hspace{10pt}(ratio 3.32)  & 3968.59 & \\
   & & \Hgamma                    & 4341.65 & $<0.9$ \\
   & & \Hbeta                     & 4862.64 & $1.9\pm0.3$ \\
   & & \OIIIall$^\ddag$           & 4960.30 & \multirow{2}{*}{$3.2\pm0.5$} \ \\
   & & \hspace{10pt}(ratio 0.335) & 5008.24 & \\
   & & \HeIL[5875]                & 5877.25 & $<0.6$ \\
   & & \OIall$^\ddag$             & 6302.05 & \multirow{2}{*}{$7.6\pm0.3$} \ \\
   & & \hspace{10pt}(ratio 3.03)  & 6363.67 & \\
   & & \Halpha+\NIIall$^\dagger$  & 6564.52 & $24.0\pm0.4$ \\
   & & \SIIall$^\dagger$          & 6725.00 & $7.6\pm0.3$ \\
   & & \HeIL[7065]                & 7067.14 & $<0.9$ \\
   & & \OIIAuall$^\ddag$          & 7321.94 & \multirow{2}{*}{$1.5\pm0.2$} \ \\
   & & \hspace{10pt}(ratio 1.92)  & 7332.21 & \\

  \midrule
  \midrule
   \multirow{11}{*}{\rotatebox[origin=c]{90}{G395H}}
   \multirow{11}{*}{\rotatebox[origin=c]{90}{\textit{ad-hoc} model \S~\ref{s.obs.ss.highres}}}
    & & Redshift                    &    ---      & $4.6348\pm0.0003$ \\
    & & Velocity dispersion $\sigma$&   [\kms]    & $400\pm35$   \\
   \cline{3-5}
    & & \multirow{2}{*}{Line(s)}   & Wavelength   & Flux \\
    & &                            & [\AA vacuum] & [\fluxcgs[-18][]] \\
   \cline{3-5}
    & & \OIall$^\ddag$              &   6302.05    &  $6.6\pm0.6$ \\ % SNR = 2.43
    & & \hspace{10pt}(ratio 3.03)   &   6363.67    &  ---         \\ % SNR = 2.43
    & & \NIIall$^\ddag$             &   6549.86    &  ---         \\ % SNR = 2.43
    & & \hspace{10pt}(ratio 0.34)   &   6585.27    &  $8.2\pm0.6$ \\ % SNR = 2.43
    & & \Halpha                     &   6564.52    & $11.6\pm0.8$ \\ % SNR = 2.43
    & & \SIIL[6716]                 &   6718.30    &  $3.7\pm0.6$ \\ % SNR = 2.43
    & & \SIIL[6731]                 &   6732.67    &  $4.2\pm0.6$ \\ % SNR = 2.43
  \bottomrule
    \end{tabular}
    \caption{Spectral measurements from the low resolution prism spectrum (top), and from the
    high-resolution G395H grating (bottom). Non detections are given as 3-\textsigma upper limits.
    The flux of lines observed in both configurations is statistically consistent.\\
    $^\dagger$ Sets of spectrally unresolved lines are treated as a single Gaussian at the reported effective wavelength.\\
    $^\ddag$ Doublets that we model as two Gaussians with fixed flux ratio; the adopted ratio (quoted in parentheses below the doublet name) is always the ratio between the blue and red line.
    }\label{t.emlines}
\end{table}

\subsection{Emission-line fitting -- grating spectrum}\label{s.obs.ss.highres}

\begin{figure*}
   \includegraphics[width=\textwidth]{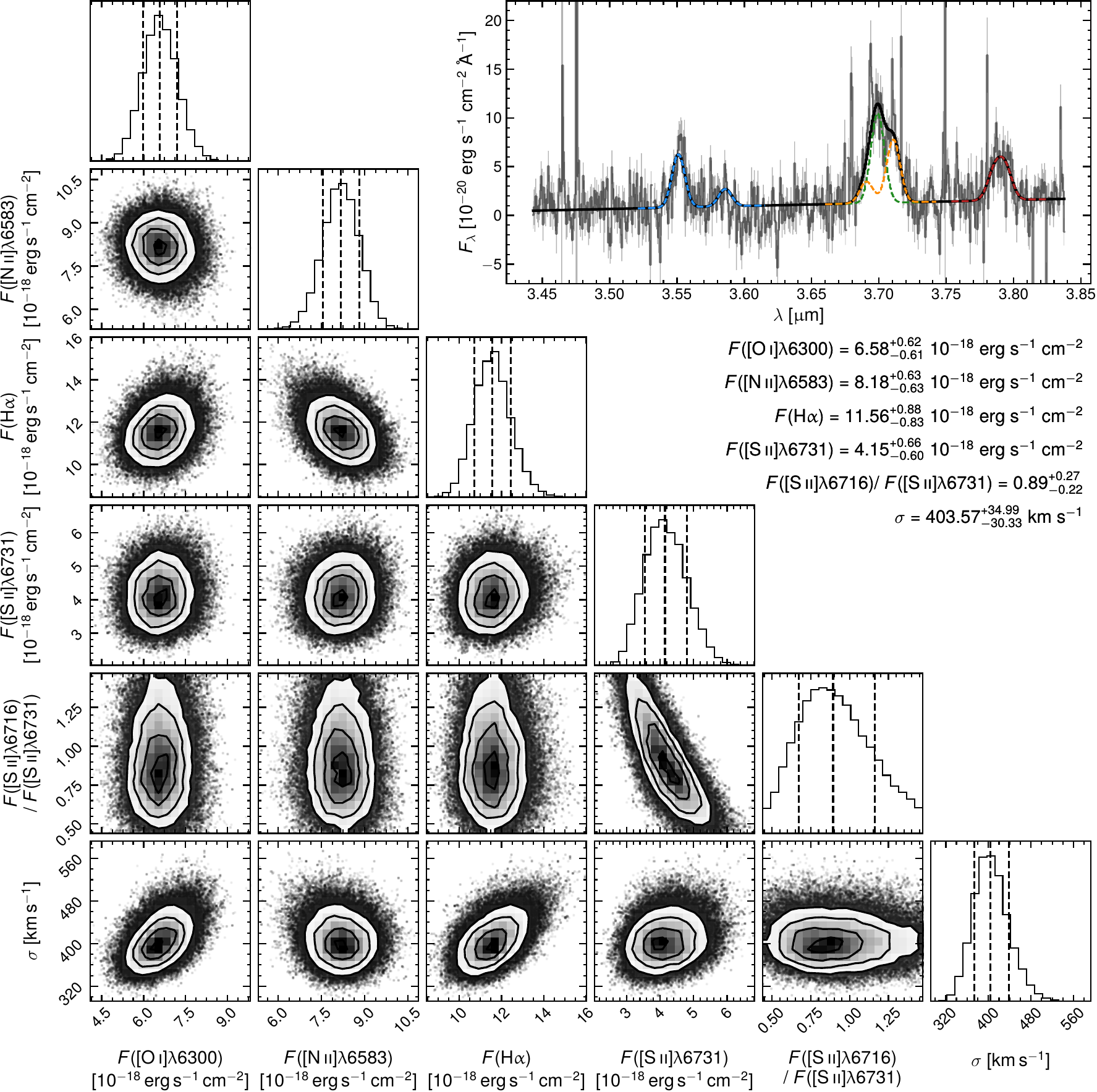}
   \caption{High spectral resolution grating observations of \target (black lines with uncertainty in grey) and fiducial model (solid black; the dashed spectra are models of the individual emission lines and doublets, from left to right \OIall, blue; \NIIall, orange; \Halpha, green; \SIIall, red).
   The corner plot reports the best-fit parameters of the emission-line model we use in the emission-line diagnostic diagrams (Fig.~\ref{f.bpt}).
   Note that the \SIIall ratio is constrained to physically permitted values, and is otherwise unconstrained by the G395H observations.
   }\label{f.highres}
\end{figure*}

In the G395H spectrum, we consider only the brightest lines, \OIall, \NIIall, \Halpha and \SIIall.
All these lines are modelled as Gaussians, with the same redshift and velocity dispersion.
The \OIall and \NIIall doublet ratios are fixed, whereas the \SIIL[6717]/\SIIL[6731] ratio is constrained between 0.44 and 1.47.
We model the background as a first-order polynomial. In total, the model has nine free parameters; the fluxes of \OIL, \Halpha, \NIIL
and \SIIL[6733]; the \SII doublet ratio; the common redshift and velocity dispersion; and two parameters for the linear background.
We adopt a Bayesian framework and estimate the posterior distribution of the parameters using a Markov-Chain Monte-Carlo (MCMC) integrator \citep[\emcee;][and references therein]{foreman-mackey+2013}.
The results are reported in Table~\ref{t.emlines}.
The data, fiducial model, and posterior distribution are shown in Fig.~\ref{f.highres}.

We find different redshifts between the prism and grating (more than 8-\textsigma significance; Table~\ref{t.emlines}). Note that a systematic difference between the NIRSpec prism and gratings has been already reported in the literature \citep{bunker+2023b}. The impact of this redshift difference on distant-dependent quantities like mass and SFR is negligible compared to the relevant uncertainties. 
We therefore adopt everywhere the redshift $z_\mathrm{spec}=4.6348$ from the grating. This measurement has formal uncertainties (inter-percentile range from the marginalised posterior) of 0.0004, corresponding to 20~\kms or half a spectral pixel, but the systematic uncertainties due to the line decomposition are probably larger.
When studying simultaneously the prism and grating spectra, we rescale the wavelength array of the prism to match $z_\mathrm{spec}$.

We note that a multi-component fit -- though certainly appropriate physically -- is not warranted by the data, due to the insufficient SNR.
We reach this conclusion because, when attempting to fit a multi-component model consisting of a set of narrow and broad emission lines, we find implausible line ratios.
Two components are clearly identified, with velocity dispersion $200\pm30$ and $600\pm100$~\kms, and with the broad component blue-shifted by $80^{+40}_{-120}$~\kms. However, the emission-line ratios of the narrow component are clearly too high -- especially if we interpret this component as a star-forming disc (we find both \OIall and \SIIall to be brighter than \Halpha). Therefore, we adopt the single-component fit as fiducial; in the next sections, we present a physically motivated multi-component approach.

\section{Shock-dominated nebular emission}\label{s.sh}

\subsection{Emission-line diagnostic diagrams}\label{s.sh.ss.bpt}

\begin{figure*}
   \includegraphics[width=0.9\textwidth]{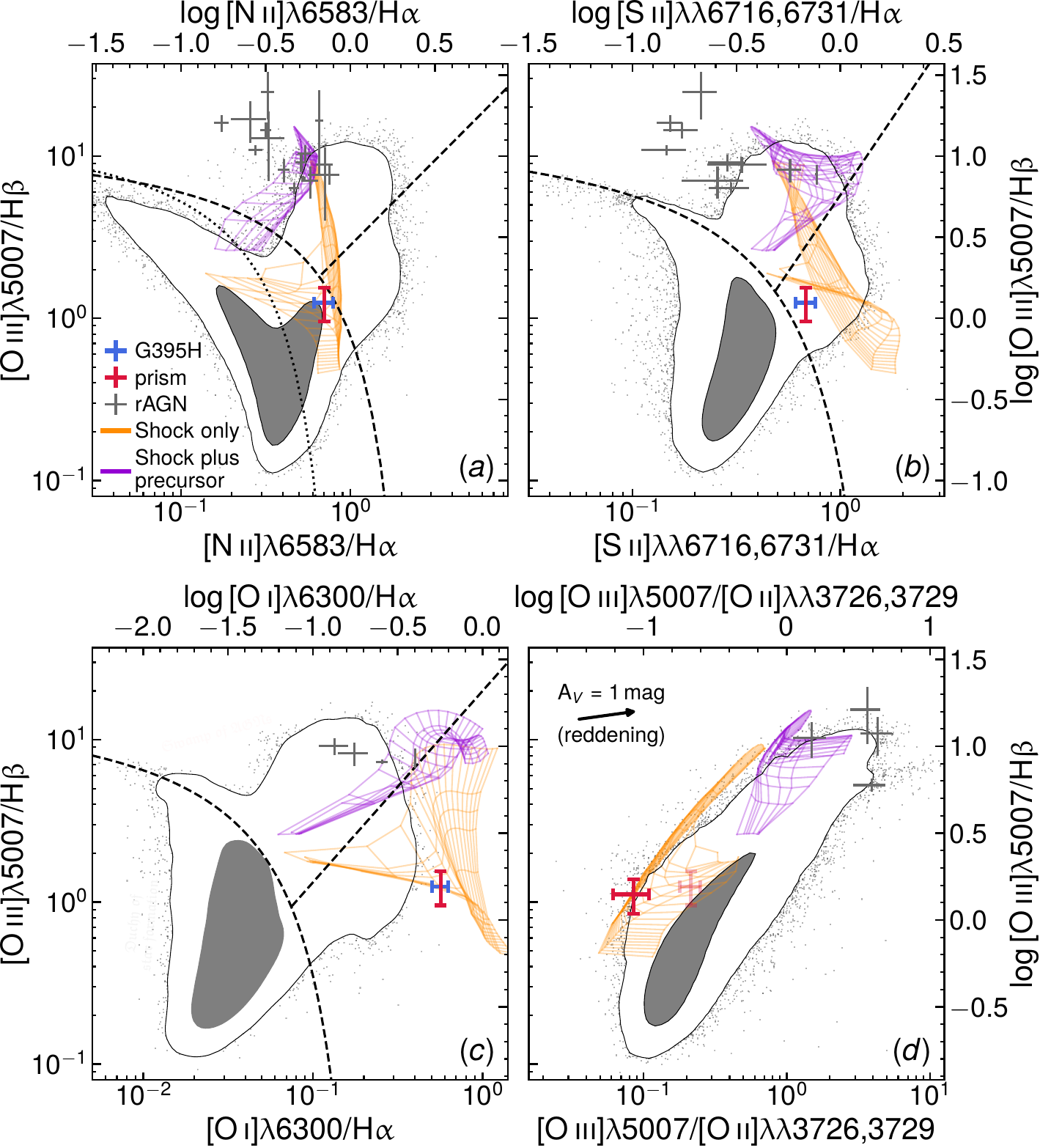}
   {\phantomsubcaption\label{f.bpt.a}
    \phantomsubcaption\label{f.bpt.b}
    \phantomsubcaption\label{f.bpt.c}
    \phantomsubcaption\label{f.bpt.d}}
   \caption{\target occupies a region of the BPT and VO diagrams \citep{baldwin+1981,veilleux+osterbrock1987} associated with strong shocks.
   Our measurements are the thick red or red and blue errobars;
   we combine measurements from both the prism (red errorbars) and G395H disperser (blue errorbars).
   In panel~\subref{f.bpt.d}, in addition to our measurement (transparent red), we also show the effect of a representative dust reddening of $A_V=1$~mag to guide the eye (solid red).
   Crosses are luminous, radio-loud AGN at redshifts $z=1.5\text{--}4$
   \citepalias[$\log P_{1.4\,\mathrm{GHz}} {[\mathrm{W\,Hz^{-1}}]} \sim 28$;][]{nesvadba+2017}. The contours and dots are local measurements from SDSS, with the
   contours enclosing the 67\textsuperscript{th} and 99\textsuperscript{th} percentiles of the data.
   The dashed and dotted demarcation lines separate regions dominated by star-formation photoionisation,
   AGNs and LIERs \citep{kewley+2001,kauffmann+2003c,kewley+2006,schawinski+2007}.
   The purple (orange) grid is a set of shock-plus-precursor (shock-only) models from
   \mappings[~v], assuming solar metallicity, density $n=1$~cm$^{-3}$ and varying
   the shock velocity and strength of the magnetic field \citep{alarie+morisset2019}.
   Our target lies closer to the shock-only models.
   }\label{f.bpt}
\end{figure*}

Combining information from the prism and G395H gratings, we can populate the classic emission-line diagnostic diagrams \citep[Fig.~\ref{f.bpt};][]{baldwin+1981,veilleux+osterbrock1987}.
The \OIIIL/\Hbeta ratio is measured from the prism spectrum, and its measurement uncertainty is highlighted in red; all other ratios were measured from the grating (highlighted in blue).
As a reference, the diagrams in Fig.~\ref{f.bpt} also illustrate two datasets; the first is the distribution of local galaxies and AGN from SDSS \cite[$z\lesssim0.1$;][the contours enclose the 68\textsuperscript{th} and 99\textsuperscript{th} percentiles of the sample; dots represent individual objects outside of the 99\textsuperscript{th}-percentile contours]{abazajian+2009}; the second dataset is the sample of bright, radio-loud AGN from \citetalias{nesvadba+2017} (shown as grey errorbars).
The dotted and dashed lines in Fig.~\ref{f.sed.a}--\subref{f.sed.c} separate three photoionisation regimes:, star-forming, AGN, and shocks/low-ionisation.
Compared to SDSS, \target occupies a sparsely populated region of the diagrams, with a rare combination of low \OIIIL/\Hbeta and high \OIL/\Halpha (panel~\subref{f.sed.c}).
However, compared to luminous radio AGN at $z=2\text{--}3$, \target is also unique; its \SIIall/\Halpha and \OIL/\Halpha are a factor of 2--3 higher than the \citetalias{nesvadba+2017} sample, whereas its \OIIIL/\OIIall and \OIIIL/\Hbeta are an order of magnitude lower.

The position of \target in Fig.~\ref{f.bpt} is consistent with local low-ionisation emission-line regions \citep[LIER;][]{heckman+1980,ho2008,belfiore+2016}, which are thought to be powered by evolved stellar populations or interstellar shocks \citep{yan+blanton2012,belfiore+2016,lee+2024}.
However, LIERs driven by stellar emission and interstellar shocks have generally low equivalent-width \Halpha emission, of the order of a few \AA \citep[e.g., `retired galaxies' in][]{cid-fernandes+2011}.
In contrast, \target has an \Halpha equivalent width EW(\Halpha)=400~\AA (measured from the prism spectrum, and correcting for \NIIall emission), more in line with other high-redshift radio AGN \citepalias{nesvadba+2017}.
In the `WHAN' diagnostic diagram, showing rest-frame EW(\Halpha) vs \NIIL/\Halpha \citep{cid-fernandes+2011}, \target occupies the top right quarter -- a position associated with shock-dominated emission \citep[e.g.,][]{drevet-mulard+2023}, yet its high EW(\Halpha) makes it an outlier.
However, unlike the sources from \citetalias{nesvadba+2017} -- which all have high line EWs -- the emission-line ratios in \target are skewed toward low-ionisation species.
Among nearby sources, the most likely spectral analogue is the \textit{IRAS} F23389+0303 ULIRG \citep[Ultra-Luminous Infra-Red Galaxy;][]{spence+2018}.
This ULIRG \citetext{also known as 4C+03.60, \citealp{gower+1967}; \href{https://skyserver.sdss.org/dr13/en/tools/explore/summary.aspx?ra=355.376275&dec=3.290678}{SDSS~J234130.31+031726.7}} is a relatively nearby object \citep[$z=0.14515\pm0.00013$;][]{spence+2018}, hosting a nearby companion \citep{kim+2002} and neutral-gas outflows \citep{veilleux+2002}. It has a $\rm{SFR}=100~\Msun~\peryr$ \citep[][comparable to our estimate for \target]{fluetsch+2021} and is a radio AGN with relatively low luminosity $P_{\rm 1.4\,GHz}=3.5\pm0.1 \times 10^{25}~\mathrm{W~Hz}^{-1}$ \citetext{up to about five times higher than \target; $S_{\rm 1.4\,GHz}=862\pm25$~mJy, \citealp{condon+1998}; we used a spectral index $\alpha=-0.81$, measured using $S_{\rm 365\,MHz}=2,563\pm63$~mJy from \citealp{douglas+1996}}.
Spectrally, F23389+0303 displays all the characteristics observed in \target, including an extremely low 
\OIIIL/\Hbeta ratio of $2.2\pm0.2$, and high emission-line ratios between the three of \OIall, \NIIall and \SIIall and \Halpha.
However, the \OIIall emission is weaker than in \target, and the auroral ratio \OIIAuall/\OIIall is considerably higher.

From the emission-line ratios in 
\target, we can measure an excitation index of 0.4~dex \citep{buttiglione+2010}. The low radio luminosity and excitation index place \target in the realm of `low excitation radio galaxies' \citep[LERG,][]{laing+1994}. Still, this classification is not without problems, because local LERGs have $EW(\OIII)<5$~\AA \citep[and typically $\sim1$~\AA;][]{best+heckman2012}, whereas for \target we find $77\pm19$~\AA. This comparison is complicated by the fact that NIRSpec MSA may be targeting an off-nuclear region in \target (Fig.~\ref{f.img}), unlike the observations of \citet{buttiglione+2010} and \citet{laing+1994}.

In Fig.~\ref{f.bpt} we also overlay a set of shock models from \citep[][see \S~\ref{s.sh.ss.shint}]{morisset+2015}, from which we selected pre-shock density $n_\mathrm{H}=1$~cm$^{-3}$, solar metallicity and the abundances of \citet{gutkin+2016}.
The two grids show shock-plus-precursor models (purple, highest 
\OIIIL/\Hbeta grids) and shock-only models (orange, lowest \OIIIL/\Hbeta grids), with the grid nodes spanning a range of magnetic field strengths $0.0001 < B < 10$~\textmu G and
shock velocities $200<v<1,000$~\kms.
Our measurements lie clearly outside the model predictions including a precursor; in contrast, the shock-only models lie reasonably close to all the observed measurements.
At the same time, all other high-redshift radio AGN present high \OIIIL/\Hbeta, and are thus more in line with high-excitation radio galaxies (HERGs).
These other galaxies agree much better with the models combining emission from the shock and the precursor and with photoionisation \citep[e.g.,][]{heckman+best2014}.

The mismatch between our measurements and the models with a precursor may be evidence for a shock that has run out of pre-shock gas.
This lack of high-ionisation emission typical of a shock precursor lends itself to the scenario of an old shock, because -- almost by definition -- a young shock is less likely to have completely swept the galaxy ISM.

Complementary evidence of a strong shock component comes from the diagnostic diagram of \citet{best+2000}, which shows \CIIIall/\CIIall vs \NeIIIL/\forbiddenEL[Ne][v]\textlambda3426 (their fig.~1). While we do not have constraints on the Ne ratio, the upper limit on the ratio \CIIIall/\CIIall places our galaxy in the lower portion of their diagram, together with other compact, radio-loud AGN \citetext{see e.g., \citealp{best+2000}, their fig.~1; or \citealp{allen+2008}, their fig.~26}.

In summary, there is evidence that a strong shock component is needed to reproduce the line ratios observed in Fig.~\ref{f.bpt} (in Appendix~\ref{a.sfonly} we also illustrate that a simple star-forming model is insufficient).
However, in the next sections we illustrate that even shocks alone cannot explain our observations.

\subsection{Shock interpolator model}\label{s.sh.ss.shint}

To efficiently generate shock models, we use the library of pre-computed shock models from the \href{https://sites.google.com/site/mexicanmillionmodels/}{database 3MdB} \citep{alarie+morisset2019}.
These models are obtained using \mappings[~v] \citep{sutherland+dopita2017}, and consist of both shock only and shock plus precursor emission \citep{allen+2008}. The models are calculated on a discrete grid spanning magnetic field strengths
$0.0001 < B < 10$~\textmu G, shock velocities $100<v<1,000$~\kms, pre-shock densities $1 < n < 10,000$~cm$^{-3}$ and metallicities  $0.0001 < Z < 0.040$ (hereafter: \textit{BnZv} space).
The chemical abundances are from \citet[][we assumed solar carbon-to-oxygen abundance ratio]{gutkin+2016}.
In particular, we use the grid \texttt{Gutkin16} and all the abundance sets from \texttt{Gutkin16\_ISM0d0001\_C1d00} to \texttt{Gutkin16\_ISM0d040\_C1d00}.
To calculate the value of the emission-line ratios at arbitrary locations in \textit{BnZv} space, we use a linear interpolator (spanning logarithmically in $B$, $n$, $Z$ and $v$).
Values outside the grid are not allowed.
Note that the shock model does not include a nebular continuum, which we discuss in \S~\ref{s.d.host}.

The shock interpolator in \textit{BnZv} (hereafter, \shinbnzv) is demonstrated in Appendix~\ref{a.shinbnzv}.
In Appendix~\ref{a.shockonly}, we use \shinbnzv and a simple dust-attenuation curve to reproduce the line ratios observed in \target. We conclude that shocks, alone, cannot simultaneously explain the high \OIIall/\Hbeta (which requires low attenuation) and the high \Halpha/\Hbeta (which requires high attenuation).
To reproduce these otherwise irreconcilable observations we need a mixture model of low- and high-attenuation spectra.
Following the analysis of \citet{moy+rocca-volmerange2002}, and the observations of \citet{drevet-mulard+2023}, we assume that this mixture consists of a combination of shocks and star-formation.

To generate simultaneously the shock emission and a galaxy SED, we incorporate \shinbnzv in the SED modelling tool \prospector (see \S~\ref{s.host}).
We use the serialisation library \texttt{pickle}, and the new class \texttt{ShockSpecModel}, derived (literally) from \prospector's \texttt{SpecModel}.
The \texttt{ShockSpecModel} class has eight new parameters:
\begin{compactitem}
    \item \texttt{shock\_type}, is either \texttt{"shock"} or \texttt{"shock+precursor"}, determines which interpolator is used
    \item \texttt{shock\_elum}, the luminosity of the \Hbeta line, in units of L$_\odot$/\mstar
    \item \texttt{shock\_eline\_sigma}, the velocity dispersion of the shock emission lines; this value is in general different from the dispersion of emission lines due to star formation, and from the shock velocity.
    \item \texttt{shock\_logB} $\in [-4, 1]$ \hfill [dex \textmu G]\phantom{aspaceaslong}
    \item \texttt{shock\_logn} $\in [0,  4]$ \hfill [dex cm$^{-3}$]\phantom{aspaceaslong}
    \item \texttt{shock\_logZ} $\in [-2.18, 0.419]$ \hfill [dex \Zsun]\phantom{aspaceaslong}
    \item \texttt{shock\_logv} $\in [2,  3]$ \hfill [dex \kms]\phantom{aspaceaslong}
    \item \texttt{shock\_dust} $\in [0, \infty]$\hfill[\texttau$_V$]\phantom{aspaceaslong}
\end{compactitem}
Here $\Zsun = 0.01542$, the value from \citet[][see \S~\ref{s.host.ss.prosp} on how we enforce consistency with the \prospector value of \Zsun]{gutkin+2016}.
The dust attenuation parameter of the shock is parametrised by the $V-$band optical depth, assuming the attenuation law of \citet{cardelli+1989}.
To take into account any wavelength-dependent mismatch between photometry and spectroscopy data, we use a calibration polynomial; this polynomial re-scales the flux of the spectrum to match the level of the photometry, because generally this kind of calibration mismatch is due to aperture losses in the spectrum.
The implications of this choice are discussed later in the article.
These polynomials are implemented following the standard method of \prospector, i.e. through the new class \texttt{ShockPolySpecModel}, which is derived from \prospector's \texttt{PolySpecModel} and from our own \texttt{ShockSpecModel}, in this order.

\section{A dusty, intermediate-mass galaxy host}\label{s.host}

\subsection{The \prospector fiducial model}\label{s.host.ss.prosp}

To model the spectro-photometric data of \target, we use v2.0 of \prospector \citep{johnson+2021}.
In addition to the photometry and prism spectrum, we also model simultaneously the high-resolution grating spectrum; fitting of multiple spectra is a feature implemented in \prospector v2.0.
We use the \texttt{ShockPolySpecModel} class, setting \texttt{shock\_type}=\texttt{"shock"}.
This choice of shock-only interpolator follows from our analysis of the emission-line ratios in \S~\ref{s.sh.ss.bpt}, due to the low \OIIIL/\Hbeta ratio and to the brightness of low-ionisation and neutral species.
We use flat, uninformative priors on all shock parameters, except for \texttt{shock\_dust}, which we set to 0, and for
\texttt{shock\_logZ}, which is tied to the nebular metallicity of the star-forming \prospector model.
The metallicity of the shock component is scaled by the ratio between the solar value in \prospector ($\Zsun = 0.0142$) and the solar value in the models of \citet[][$\Zsun = 0.01542$]{gutkin+2016}, such that the absolute metallicity is consistent. We make no attempt to match the detailed chemical abundances.

We use a non-parametric star-formation history (SFH) with continuity priors \citep{leja+2019}. We rely on nine time bins, with the two most recent bins having duration 30 and 100~Myr, followed by seven logarithmically spaced bins between 130~Myr and redshift $z=20$ (no stars are formed earlier).
We use a flat prior on stellar mass and metallicity (in log space), the \citet{kriek+conroy2013} attenuation law, with extra attenuation towards birth clouds, following \citet{charlot+fall2000}.
Overall, the parameters of the host galaxy follow the setup of \citet[][hereafter: \citetalias{tacchella+2022a}]{tacchella+2022a}, including coupling ongoing star formation to nebular emission. A summary of the model parameters and their prior probability distributions is presented in Table~\ref{t.prosp}.

To adjust the weight of photometry and spectroscopy, we use the noise `jitter' term (on the spectrum only), which can scale the input noise vector by a uniform factor (with flat prior between 0.5 and 2).
While \prospector can model outliers with a mixture model, the emission-line ratios of \target are so extreme that we risk masking some lines. For this reason, we mask spectral pixels identified as outliers in the input spectrum, and do not use any outlier model at runtime.
The posterior probability is estimated using nested sampling \citetext{\citealp{skilling2004}, using the library \textsc{dynesty}; \citealp{speagle2020}}, but we checked that our results do not change using the MCMC integrator \textsc{emcee}.
We provide an elementary test of this combination of data and model in Appendix~\ref{a.recovery}, where we use a set of mock observations to show that the model is capable of recovering the input parameters, albeit with some bias in $B_\mathrm{sh}$ and $n_\mathrm{sh}$.

\begin{table*}
    \begin{center}
    \caption{Summary of the parameters, prior probabilities and posterior probabilities of the fiducial \prospector model (see also Fig.~\ref{f.sed}). 
    }\label{t.prosp}
    \setlength{\tabcolsep}{4pt}
    \begin{tabular}{llcllc}
  \toprule
   & Parameter & Free & Description & Prior & Posterior \\
   & (1)       & (2)  & (3)         & (4)   & (5) \\
   \midrule
   \multirow{11}{*}{\rotatebox[origin=c]{90}{Star-forming component}}
   & $z_\mathrm{obs}$ & Y & redshift & $\mathcal{N}(z_\mathrm{spec}, 0.001)$ & $4.6317^{+0.0002}_{-0.0002}$ \\
   & $\log \mstar [\Msun]$ & Y & total stellar mass formed & $\mathcal{U}(7, 13)$ & $10.16^{+0.12}_{-0.09}$\\
   & $\log Z [\Zsun]$ & Y & stellar metallicity & $\mathcal{U}(-2, 0.19)$ & $-0.33^{+0.12}_{-0.13}$ \\
   & $\log \mathrm{SFR}$ ratios & Y & ratio of the $\log \mathrm{SFR}$ between adjacent bins of the non-parametric SFH & $\mathcal{T}(0, 0.3, 2)$ & --- \\
   & $\sigma_\star \; [\kms]$ & Y & stellar intrinsic velocity dispersion & $\mathcal{U}(0, 450)$ & $231^{+40}_{-37}$ \\
   & $n$ & Y & power-law modifier of the \citet{kriek+conroy2013} dust curve \citepalias[][their eq.~5]{tacchella+2022a} & $\mathcal{U}(-1.0,0.4)$ & $0.03^{+0.06}_{-0.07}$ \\
   & $\tau_V$ & Y & optical depth of the diffuse dust \citepalias[][their eq.~5]{tacchella+2022a} & $\mathcal{G}(0.3,1;0,2)$ & $0.74^{+0.09}_{-0.09}$ \\
   & $\mu$ & Y & ratio between the optical depth of the birth clouds and $\tau_V$ \citepalias[][their eq.~4]{tacchella+2022a} & $\mathcal{U}(-1.0,0.4)$ & $1.08^{+0.18}_{-0.18}$ \\
   & $\sigma_\mathrm{gas} \; [\kms]$ & Y & intrinsic velocity dispersion of the star-forming gas & $\mathcal{U}(30,300)$ & $231^{+32}_{-28}$ \\
   & $\log Z_\mathrm{gas} [\Zsun]$ & Y & metallicity of the star-forming gas & $\mathcal{U}(-2, 0.19)$ & $0.03^{+0.02}_{-0.02}$ \\
   & $\log U$ & Y & ionisation parameter of the star-forming gas & $\mathcal{U}(-4, -1)$ & $-3.39^{+0.19}_{-0.11}$\\
   \midrule
   \multirow{6}{*}{\rotatebox[origin=c]{90}{Shock component}}
   & $L(\Hbeta)_\mathrm{sh} [\mathrm{L}_\odot \mstar^{-1}]^\ddag$ & Y & luminosity of the \Hbeta line in the shock component & $\mathcal{U}(0,10)$ & --- \\
   & $\sigma_\mathrm{sh} \; [\kms]$ & Y & intrinsic velocity dispersion for the shocked gas & $\mathcal{U}(100,1,000)$ & $478^{+15}_{-18}$ \\
   & $\log B_\mathrm{sh} \; [\text{\textmu} G]$ & Y & strength of the magnetic field & $\mathcal{U}(-4,1)$ & $0.63^{+0.12}_{-0.21}$ \\
   & $\log v_\mathrm{sh} \; [\kms]$ & Y & shock velocity & $\mathcal{U}(2, 3)$ & $2.88^{+0.05}_{-0.09}$ \\
   & $\log n_\mathrm{sh} \; [\mathrm{cm}^{-3}]$ & Y & pre-shock density & $\mathcal{U}(0, 4)$ & $1.19^{+0.10}_{-0.11}$ \\
   & $\log Z_\mathrm{sh} [\Zsun]$ & N & shock metallicity & tied to $\log Z_\mathrm{gas}$ & --- \\
   & $\tau_{V,\mathrm{sh}}$ & N & shock optical depth, assuming the \citet{cardelli+1989}
   attenuation law & 0 & --- \\
   \midrule
   \multirow{3}{*}{\rotatebox[origin=c]{90}{Other}}
   & $j_\mathrm{spec}$ & Y & multiplicative noise inflation term for spectrum & $\mathcal{U}(0.5,2)$ & $1.0^{+0.3}_{-0.3}$ \\
   & \fHbetash & N & flux ratio between the shock component and the total for \Hbeta & --- & $0.54^{+0.02}_{-0.03}$\\
   & $\log SFR_{10} [\Msun \, \peryr]$ & N & star-formation rate averaged over the last 10~Myr & --- & $1.86^{+0.06}_{-0.08}$ \\
   & $\log SFR_{100} [\Msun \, \peryr]$ & N & star-formation rate averaged over the last 100~Myr & --- & $1.83^{+0.08}_{-0.13}$ \\
  \bottomrule
  \end{tabular}
  \end{center}
(1) Parameter name and units (where applicable). (2) Only parameters marked with `Y' are optimised by \prospector; parameters marked with `N' are either tied to other parameters (see Column~4), or are calculated after the fit from the posterior distribution (in this case, Column~4 is empty). (3) Parameter description. (4) Parameter prior probability distribution; $\mathcal{N}(\mu, \sigma)$ is the normal distribution with mean $\mu$ and dispersion $\sigma$; $\mathcal{U}(a, b)$ is the uniform distribution between $a$ and $b$; $\mathcal{T}(\mu, \sigma, \nu)$ is the Student's $t$ distribution with mean $\mu$, dispersion $\sigma$ and $\nu$ degrees of freedom; $\mathcal{G}(\mu, \sigma, a, b)$ is the normal distribution with mean $\mu$ and dispersion $\sigma$, truncated between $a$ and $b$.
(5) Median and 16\textsuperscript{th}--84\textsuperscript{th} percentile range of the marginalised posterior distribution; for some nuisance parameters we do not present the posterior statistics (e.g., log SFR ratios). $^\ddag$ Units of solar luminosity per total stellar mass formed in \Msun.
\end{table*}

\subsection{\prospector results}\label{s.host.ss.results}

The fiducial \prospector model is shown in Fig.~\ref{f.sed}, with the marginalised posterior distribution for a selected subset of the parameters (a more extensive list of parameters and their posterior probabilities is listed in Table~\ref{t.prosp}).
The target has a relatively low value of \mstar ($\log \mstar/\Msun = 10.16^{+0.12}_{-0.09}$ Table~\ref{t.prosp}), smaller than the value reported in \citet[][$\log \mstar/\Msun = 10.4$~dex]{stefanon+2017} which was based on photometry alone.
The redshift difference \citetext{\citealp{stefanon+2017} used a photometric redshift $z=4.2$} makes the discrepancy even worse (by 25~per cent).
The main driver of this discrepancy is information from the spectrum, not including the emission from the shock; indeed, the star-forming model without shocks
finds very similar \mstar to the fiducial run with shocks (Appendix~\ref{a.sfonly}).
It is only by removing spectroscopy that the \prospector model is able to reproduce the higher \mstar values reported in the literature.
The fiducial \mstar value is remarkably small, considering that the knee of the galaxy mass function at redshifts $z=4\text{--}5$ is of order $3\times 10^{10}$~\Msun \citep[for both star-forming and quiescent galaxies;][]{weaver+2023}.

The SFH of this fiducial model is traced by the solid red line in Fig.~\ref{f.sed.f}; the shaded pink region is the uncertainty, the blue line is the locus of the star-forming sequence for galaxies with mass equal to the stellar mass formed at each epoch \citep[][their equation~10]{popesso+2023}, and the horizontal dashed red line is the quiescent threshold defined by $\mathrm{sSFR} < 0.2 / t_U(z_\mathrm{spec})$, where $t_U(z)$ is the age of the Universe at redshift $z$ \citep[e.g.,][]{franx+2008,pacifici+2016}.

Using more bins at recent times, or fewer bins at earlier times does not affect our results significantly. For instance, adding a time bin between look-back times of 0 and 10~Myr, most quantities remain within 1~\textsigma from their value in the fiducial run; the exceptions are $\log n$ and $\log B_\mathrm{sh}$, which both increase by 1.5~\textsigma, and $\mu$, which decreases by 2~\textsigma from 1.41 to 1.03 (i.e., it reverts to the mean of the truncated-Gaussian prior, see Table~\ref{t.prosp}).
Similarly, we find little differences when changing the prior on the SFH to be more bursty; increasing the dispersion of the Student's $t$ priors from 0.3 to 1, we find that the galaxy had higher SFRs at recent times (100--300~Myr), and an overall shorter SFH, resulting in roughly the same \mstar.
We also tested a model with a delayed exponential SFH plus a single, instantaneous burst (dotted grey line in Fig.~\ref{f.sed.f}); the model finds a much higher value of \fHbetash (0.9), but the attenuation is higher and overall \mstar, the SFR and the shock parameters are almost unchanged.
The relatively robust SFH derives from the shape of the continuum, including the weak Balmer break and mildly rising UV continuum.
The presence of a Balmer break argues against an AGN-dominated continuum. Overall, \prospector finds a SFH that is strongly rising at recent times, placing the galaxy on or even above the star-forming sequence (Fig.~\ref{f.sed.f}).

\begin{table}
    \centering
    \setlength{\tabcolsep}{4pt}
    Emission-line fluxes from the fiducial \prospector model (Table~\ref{t.prosp} and Fig.~\ref{f.sed}).
    \begin{tabular}{lccc}
  \toprule
    \multirow{2}{*}{Line(s)}  & Flux (SF)         & Flux (Shock) & Ratio \\
                              & \multicolumn{2}{c}{[\fluxcgs[-18][]]} & Shock to total \\
  \bottomrule
    \OIIL[3726]  &  $6.0\pm0.8$ & $32.1\pm2.2$ & $0.84$ \\
    \OIIL[3729]  &  $8.1\pm1.1$ & $17.3\pm1.4$ & $0.68$ \\
    \Hbeta       &  $6.4\pm0.6$ & $7.4\pm0.4$ & $0.54$ \\
    \OIIIL[5007] &  $3.0\pm1.3$ & $6.7\pm1.2$ & $0.69$ \\
    \OIL[6300]   &  $1.5\pm0.3$ & $23.2\pm1.1$ & $0.94$ \\
    \Halpha      & $30.3\pm2.1$ & $22.5\pm1.3$ & $0.43$ \\
    \NIIL[6583]  & $16.5\pm1.6$ & $16.7\pm2.0$ & $0.50$ \\
    \SIIL[6716]  & $10.1\pm1.8$ & $6.0\pm1.0$ & $0.37$ \\
    \SIIL[6731]  &  $7.8\pm1.4$ & $8.9\pm1.7$ & $0.53$ \\
    \OIIAuall    &  $0.4\pm0.1$ & $5.5\pm0.5$ & $0.92$ \\
  \bottomrule
    \end{tabular}
    \caption{The model separates emission from shocks and from star-formation photo-ionisation. We also report the fractional contribution of shocks. The shock fluxes assume no dust attenuation (Table~\ref{t.prosp}). All fluxes are up-scaled to match the normalisation of the photometry; for this reason, all these fluxes are $\approx 5$ times brighter than the measurements (Table~\ref{t.emlines}).
    }\label{t.emlines.prosp}
\end{table}

\subsection{Modelling pitfalls}\label{s.host.ss.pitfalls}

\begin{figure*}
   \includegraphics[width=\textwidth]{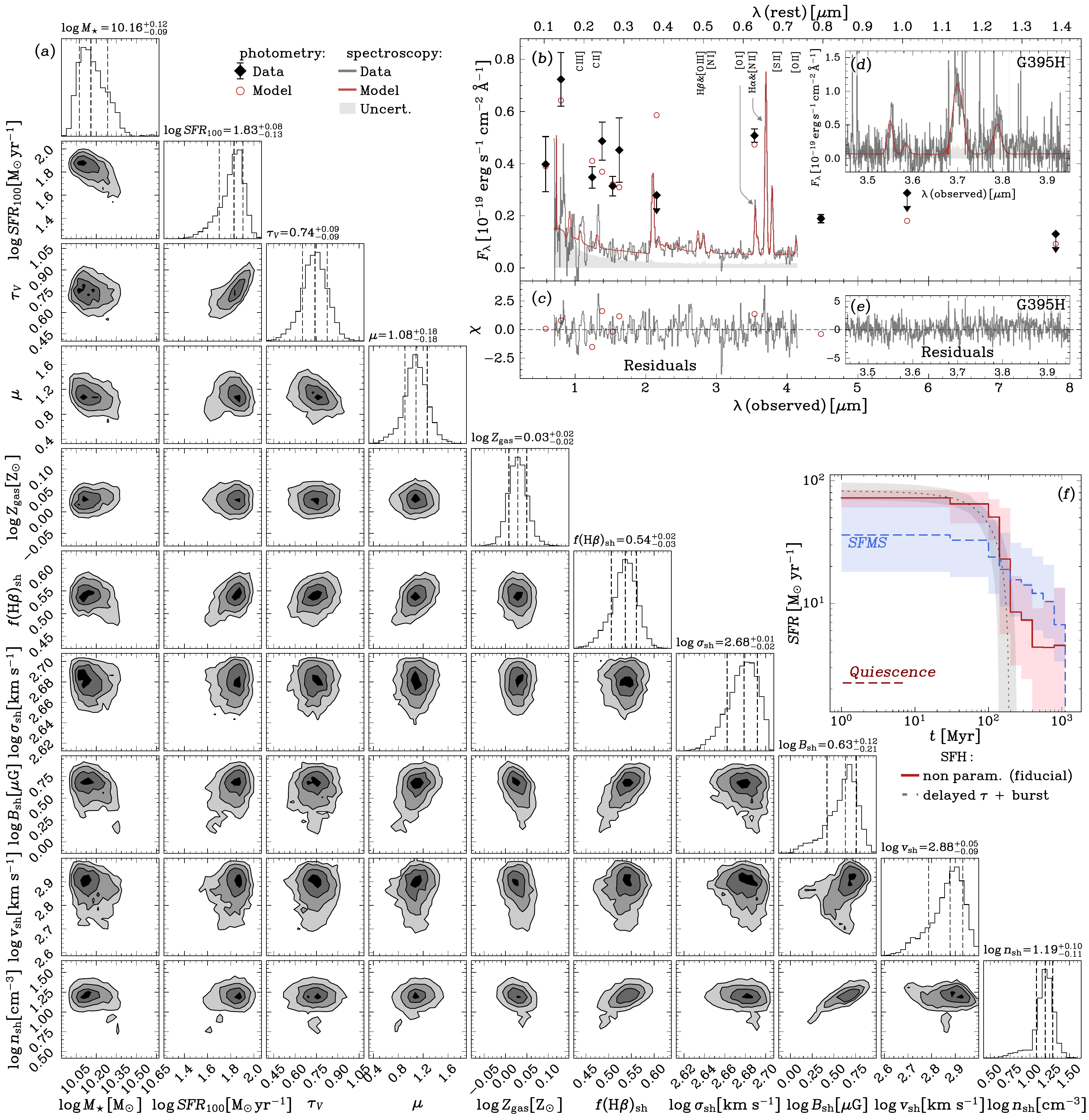}
   %python3 nice_plot_ulema_logshock.py 2018003848_nonp_r100_r2700_constr_all_lint_24Feb26-16.45.h5
   {\phantomsubcaption\label{f.sed.a}
    \phantomsubcaption\label{f.sed.b}
    \phantomsubcaption\label{f.sed.c}
    \phantomsubcaption\label{f.sed.d}
    \phantomsubcaption\label{f.sed.e}
    \phantomsubcaption\label{f.sed.f}}
   \caption{Posterior distribution for a selection of parameters of the spectro-photometric model of \prospector, alongside a comparison between data and model (panels~\subref{f.sed.b} and \subref{f.sed.d}, for prism and G395H, respectively), the normalised residuals $\chi$
   (panels~\subref{f.sed.c} and \subref{f.sed.e}), and the reconstructed star-formation history (panel~\subref{f.sed.f}).
   We find that \fHbetash, the \Hbeta flux fraction due to the shock, is equally split between
   shock emission and star-formation photoionisation, implying that
   a shock-only or star-formation-only model may not capture the properties
   of this target. The gas metallicity is found to be approximately solar.
   The overall flux scaling is set by the photometry (circles and errorbars).
   The presence of a Balmer/4000-\AA break argues against an AGN-dominated continuum, and in favour of a stellar origin.
   The SFH is dominated by a high SFR in the most recent 200~Myr, which places the galaxy on or above the star-forming sequence \citep[dashed blue line][]{popesso+2023} and clearly above the quiescence threshold (dashed red horizontal line). Using a non-parametric SFH (red solid line, fiducial) or a delayed-$\tau$ model plus burst (grey dotted line) gives very similar results. We find no evidence of quenching, suggesting that radio-AGN feedback is not coupled with the star-forming gas.
   See also Table~\ref{t.prosp}.}\label{f.sed}
\end{figure*}

In Fig.~\ref{f.sed.b} we compare the maximum a-posteriori prospector model with the photometry (diamonds) and prism spectroscopy (black line).
We find several discrepancies, which we discuss in the following.

There is a wavelength offset of $\approx 1\text{--}2$ pixels at the location of \CIIall.
A strong auroral \CIIall is a signature of shocks \citep[e.g.,][]{best+2000}, where this line can be comparable or even brighter than \Hbeta \citep{allen+2008}, so its detection in \target would not be surprising, and would indicate an overall low dust attenuation.
The wavelength mismatch could indicate a problem in the wavelength calibration of the prism spectrum, similar to what reported by other works \citep{deugenio+2024}. Alternatively, the observed spectral feature is not \CIIall, but an artefact; deeper observations would certainly clarify this point.

Another mismatch between data and observations is the over-predicted $K_s$-band flux, with the model value greatly exceeding the WIRCAM upper limit (panel~\ref{f.sed.b}).
This mismatch is certainly due to the strong \OIIall in the spectrum, and its much lower equivalent width in the photometry; indeed, clear indications of this spatial dependence are visible in the 2-d spectrum, where the \OIIall is barely detected beyond the central shutter (Fig.~\ref{f.sed.b}).
This indicates that the shock is spatially compact, relative to the size of the galaxy -- in agreement with the expectation for low-power radio AGN.

The model is biased to higher \Hbeta and lower \OIIAuall, which is explained in part by the low weight these spectral features carry in the posterior probability (6 and 7-\textsigma detection, respectively).
Lowering artificially the noise around \Hbeta--\OIIIall and \OIIAuall (by a factor of ten), the model predicts these lines correctly, but is unable to match the strong \OIIall emission.
To improve the fit around \Hbeta and \OIIAuall, we require introducing two more free parameters, the metallicity of the shocked gas (independent from the metallicity of the star forming gas), and the shock dust attenuation.
If we add only the shock attenuation, we find $\tau_{V,\mathrm{sh}}=3.4$ ($A_{V,\mathrm{sh}}=3.7$~mag), a value that is high, but comparable to radio-loud AGN at high redshift \citepalias{nesvadba+2017}. Crucially, the model predicts correctly \OIIAuall, but fails to predict both \Hbeta and \OIIall. The resulting stellar mass is 0.2~dex lower than in the fiducial case.
If we add only the shock metallicity, the model correctly predicts \OIIall, but fails to predict \OIIAuall; the resulting posterior has 0.2-dex higher \mstar, and different metallicities for the star-forming and shock gas ($\log\,Z_\mathrm{gas}/\Zsun = -0.38$ and $\log\,Z_\mathrm{sh}/\Zsun = 0.30$, respectively).
These two alternative models give unrealistic values of either $A_{V,\mathrm{sh}}$ or $Z_\mathrm{sh}$.
As a third possibility, adding both $\tau_{V,\mathrm{sh}}$ and a free $Z_\mathrm{sh}$, instead, provides a physically plausible solution.
The fiducial model parameters remain mostly unchanged, but the galaxy dust attenuation becomes stronger -- particularly for the birth-cloud component, where it reaches an optical depth of $\approx 2$.
The shock dust attenuation, in contrast, is relatively low, at $\tau_{V,\mathrm{sh}}\approx0.9$.
This dust attenuation requires higher intrinsic \OIIall flux, which the model is able to reproduce with shock metallicity $\log\,Z_\mathrm{sh}/\Zsun=0.03$; at the same time, the requirements on the star-forming gas are lowered to sub-solar values of $\log\,Z_\mathrm{gas}/\Zsun=-1.4$.
This expanded model reproduces the data slightly better than the fiducial model, but the reduced $\chi^2$ is indistinguishable from the fiducial model (to within less than 1~per cent), so we have no reason to prefer one model over the other.

\section{Discussion}\label{s.disc}

\subsection{Comparison with photoionisation equilibrium models}\label{s.d.temden}

Analysing the physical properties of line-emitting plasma is often done under the assumption that all nebular emission arises from a region with uniform density, temperature and chemical abundance, in both thermal and ionisation equilibrium.
In contrast, the shock models we use encompass a diverse range of densities, temperatures and ionisation conditions \citep{allen+2008,alarie+morisset2019}.
It is therefore interesting to assess how our results compare to the common approach of photoionisation equilibrium.

Using the emission-line measurements of Table~\ref{t.emlines}, we would reach implausibly high metallicity.
This is driven by the high Balmer decrement of $\approx 6$; interpreting this ratio under the usual assumptions of Case~B recombination and standard density and temperature \citep[intrinsic $\Halpha/\Hbeta = 2.86$,][]{osterbrock+ferland2006}, we require a dust attenuation of $A_V \approx 2.4$~mag. Even assuming higher values of the intrinsic ratio \citep[$\Halpha/\Hbeta=3$; e.g.,][]{tacchella+2022b}, the dust attenuation required would still be high ($A_V = 2.2$~mag). Such high attenuation increases the metallicity estimate in two ways \citetext{we assumed the \citealp{cardelli+1989} attenuation law}; 
on one hand, high attenuation lowers the ratio between auroral and strong \OII from 0.14 to 0.02, which in turn lowers the temperature to $T_\mathrm{e}\approx8,000$~K -- a low temperature increases the recombination rate of hydrogen and reduces the emissivity of collisionally excited metal lines. On the other hand, high attenuation increases the \OIIall/\Hbeta ratio from the observed value of $\approx 6$ to 13. With these values and a mean density of 1,000~cm$^{-3}$ (from \SII), we get a O$^+$/H$^+$ abundance of 0.0016. Neglecting higher ionisation states, this corresponds to $12+\log(\mathrm{O/H})=9.2$~dex, or [O/H]=0.51~dex with the solar metallicity from \citet{gutkin+2016}.
This exercise illustrates the perils of assuming the same physical properties and dust attenuation for all emission lines. For our case, part of the Balmer emission comes from higher-ionisation regions than the \OII and \SII lines, so estimating the dust attenuation from the Balmer decrement and applying the resulting correction to all emission lines results in implausible gas properties, like super-solar metallicity at $z=4.6$.

\begin{figure}
  \includegraphics[width=\columnwidth]{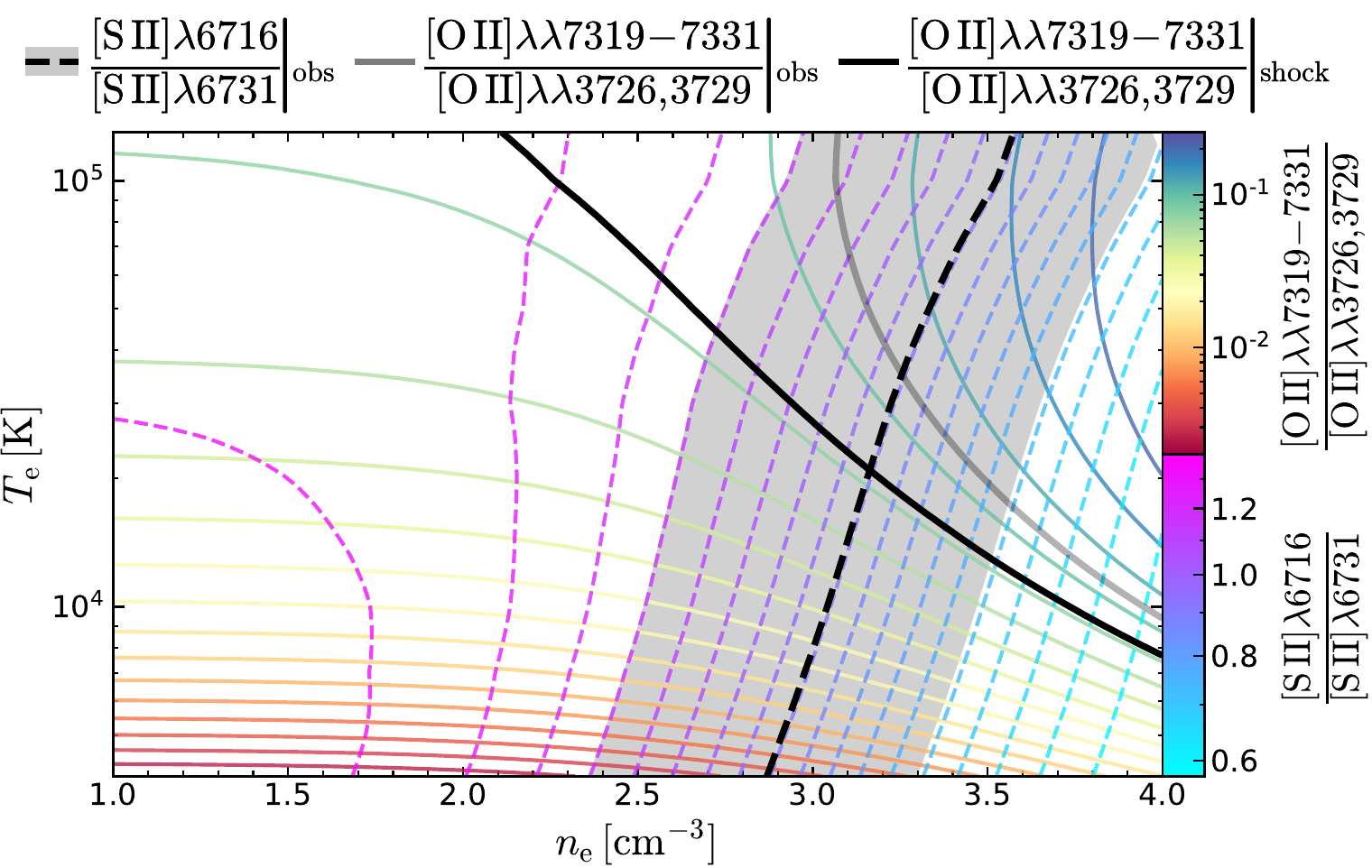}
  \caption{Predicted line ratios of \SII and \OII as a function of density and temperature, for an equilibrium model. The dashed (solid) coloured lines show the predicted ratios of \SII (\OII), according to the colourbar; the observed \SII ratio is the thick dashed line (with the grey region encompassing the uncertainties); the observed \OII ratio is the thick solid grey line, while the thick solid black line is the \OII ratio from the \prospector shock model.
  Emissivities were calculated using \pyneb.
  }\label{f.temden}
\end{figure}

Limiting ourselves to the shock emission, and still under the assumptions of equilibrium, we use \pyneb \citep{luridiana+2015} to calculate the emissivity of S$^+$ and O$^+$ on the density--temperature plane, and to compare the predicted emission-line ratios to the values measured in \target.
In Fig.~\ref{f.temden}, the grid is rendered by thin dashed and solid lines, coloured respectively according to the ratios of the \SII and \OII lines; the observed ratios are the thick lines, either dashed (for \SII) or solid (\OII).
The measurement uncertainties on the \SII ratio are encompassed by the grey-shaded region; as we pointed out, this measurement is not the most constraining, due to the combination of broad line profile and low SNR (Fig.~\ref{f.highres}).
The thick solid black line is the \OII ratio for the shock component of the best-fit \prospector model; due to the negligible contribution of the star-forming component to the \OIIAuall flux, the shock component should have even higher red-to-blue ratio than the empirical measurement. However, as we have pointed out, the fiducial model under-predicts \OIIAuall, resulting in a lower ratio than what is measured in the prism data (0.10 vs 0.14).
The intersection of the solid black line with the shaded region from \SII gives us $\log n_\mathrm{e} [\mathrm{cm}^{-3}] = 3.2$ and $T_\mathrm{e}=20,000$~K, while the limits of the grey shaded region give $2.7 < \log n_\mathrm{e} [\mathrm{cm}^{-3}] < 3.5$ and $44,000 > T_\mathrm{e} > 13,000$~K.

Based on the equilibrium values of $n_\mathrm{e}$ and $T_\mathrm{e}$,  we obtain a metallicity $12 + \log (\mathrm{O/H}) = 7.59$~dex, or 10~per cent solar.
The large discrepancy (one dex) between this value and the value inferred from the \prospector model with shocks underscores the limitations of the approach proposed in this section.

\subsection{Dust attenuation and metallicity}

While our models find solar metallicity, there are several caveats to this result.
First, the fiducial model does not reproduce perfectly the observed \Hbeta flux, which is over-estimated by approximately 30~per cent; this suggests an even higher metallicity.
Second, we have assumed a uniform gas-phase metallicity for both star-forming regions and the shocked gas. In reality, the shocked gas has a different spatial distribution compared to the star-forming gas (as evidenced by the spatial variation in the emission-line ratios, Fig.~\ref{f.img.b}). Therefore, it is possible that the shock-driven emission arises from a different region of the galaxy, possibly a small, highly enriched central zone, near to the galaxy SMBH.
It is possible -- though still unclear -- that these regions undergo fast-track chemical enrichment \citep[e.g.,][]{hamann+ferland1992,baldwin+2003,matsuoka+2017,ji+2024a}.
The alternative interpretation, i.e. that the entire ISM had solar metallicity, runs against the expectations of the mass--metallicity relation at $z\approx4$ \citep[e.g.,][]{curti+2023b,nakajima+2023}, which for a galaxy with the mass of \target predicts a metallicity of 0.25--0.3~\Zsun; admittedly, current sample sizes at $z>4$ are still small, so the scatter about the mass--metallicity relation is not known precisely.

Our data do not prefer the dual-metallicity over the single-metallicity model, but the dual-metallicity model gives a combination of solar metallicity for the shocked gas, and an extremely low gas metallicity of $\log\,Z_\mathrm{gas}/\Zsun = -1.4$ for the star-forming gas.
This value is roughly ten times lower than the prediction of the mass-metallicity relation, so we can conclude that the dual-metallicity model does not alleviate the tension between our galaxy and the metallicity scaling relations. The SNR of our data does not allow to constrain the metallicity of both regions.

Note that the dual-metallicity model also requires significant dust attenuation in the shock component -- although this amount is significantly less than what found for the star-forming gas ($\tau_{V,\mathrm{sh}}=0.9$ vs $\tau_V = 2$~mag for the star-forming gas).
Dust is not expected to survive in shocks, due to a number of physical processes \citep[e.g.,][]{draine+salpeter1979,dwek+1996,jones+1996,dwek+1996,pineaudesforets+flower1997,perna+lazzati2002}.
This is particularly relevant for fast shocks like the case in hand, based on the inferred shock velocity of over 600~\kms (Table~\ref{t.prosp}).
A modest amount of attenuation could come from a foreground dust screen, although we note that there are local cases of spatially resolved dust attenuation in jet-driven shocks \citep{drevet-mulard+2023}, i.e., some dust could be still embedded within or very near to the shocked medium.
The presence of blue emission in \hst imaging around the \Lyalpha drop (Fig.~\ref{f.img.a}) -- if it arises from the galaxy and not from an interloper -- suggests a complex dust structure with incomplete covering.
Dust destruction could also play a role in explaining the high nebular metallicity. 
Typical dust-to-metal mass ratios are in the range 0.1--0.3 \citep{devis+2019}; a significant suppression of this ratio, due to the shock destroying dust, could therefore transfer metals from the solid to the gas phase, hence increasing the gas-phase metallicity by up to an order of magnitude.

\subsection{Host-galaxy properties}\label{s.d.host}

One of the goals of our simultaneous model of the host-galaxy SED and shocks-driven emission is to quantify the impact of neglecting shock emission, and degeneracies between the model parameters describing the host and the shocks.
By comparing the fiducial model to the no-shocks model (Figs.~\ref{f.sed} and~\ref{f.a.prosp}), we find that neglecting the shock emission has a negligible impact on \mstar and SFR.
This is surprising, given the high equivalent width of the nebular emission lines.
Inspecting the posterior probability distribution from Fig.~\ref{f.sed.a}, we find that most of the shock and galaxy parameters are fairly decoupled, while the usual degeneracies within each parameter subset are readily recovered (e.g., \mstar--SFR, SFR--$\tau_V$; or $B_\mathrm{sh}$--$n_\mathrm{sh}$).
However, we also find significant degeneracy between \fHbetash and SFR, $\tau_V$ and $\mu$ (the extra attenuation toward stellar birth clouds).
Using a partial-correlations approach \citep[e.g.,][]{bait+2017,vallat2018,bluck+2019}, we identify the correlation between \fHbetash and $\mu$ as the most significant, with a Spearman rank correlation coefficient $\rho = 0.7$ after controlling for $\tau_V$ and SFR.
The positive correlations between \fHbetash and $\mathrm{SFR}_{10}$ and $\mathrm{SFR}_{100}$ are less strong ($\rho = 0.39$ and 0.44, respectively).
Naively, one would expect an anti-correlation between \fHbetash and SFR -- particularly with $\mathrm{SFR}_{10}$; this is because the \Hbeta flux is constrained by the data, so a higher \fHbetash value decreases the fraction of \Hbeta flux available for star formation.
What is happening, however, is that increasing \fHbetash also increases the attenuation $\mu$ towards birth clouds; this means that the observed fraction \fHbetash increases, the total dust-corrected \Hbeta flux also increases, thus increasing $\mathrm{SFR}_{10}$. 
This is underscored by the fact that the correlations between \fHbetash and SFR are significantly reduced or even reversed after controlling for $\mu$ (partial correlation coefficient $\rho = -0.50$ and $0.26$ for $\mathrm{SFR}_{10}$ and $\mathrm{SFR}_{100}$, respectively).
Ultimately, it is this correlation between \fHbetash and $\mu$ that drives the robustness of \mstar and SFR against including or excluding shocks.

The correlation between \fHbetash and $\mu$ itself is harder to track down, but a plausible explanation is that solutions with high \fHbetash and low $\mu$ cannot explain the high observed \Halpha flux.
If this explanation is correct, then galaxies with similar dust attenuation between the shock- and star-forming emission may have strong degeneracies between \fHbetash and SFR$_{10}$ and \mstar.

A significant limitation of our approach is that we do not include the shock nebular continuum.
For low resolution spectroscopy, a strong nebular continuum is significantly degenerate with stellar age, and a higher nebular-continuum fraction in the UV can be hidden by a strong Balmer break in stellar atmospheres.
Indeed, there are renowned examples of radio galaxies with strong nebular continuum, with UV fractions as high as 50--80~per cent \citep[e.g., 3C~368][]{dickson+1995,stockton+1996}, and these have high \OIIall/\Hbeta ratios, similar to \target \citep{inskip+2002}.
However, in these cases, the nebular continuum is readily seen in the spectrum via the Balmer jump, which is not the case in \target (Fig.~\ref{f.img.c}; although, due to low SNR and spectral resolution, we cannot fully rule out a combination of shock-driven Balmer jump and stellar Balmer break).
Moreover, a strong nebular continuum is normally associated with strong recombination lines like \Hbeta, while our target has remarkably weak \Hbeta.
Regardless, generalising the present model to include the shock continuum would require a different approach than what used for \shinbnzv.

The stellar mass and SFH we infer are remarkably robust against different modelling choices, from including/removing shock emission, to changing the SFH priors and parametrisations (Fig.~\ref{f.sed.f}).
The SFH parametrisation should carry a systematic error of 0.2--0.3~dex \citep[e.g.,][]{leja+2019a,carnall+2019,lower+2020}, but in our case the non-parametric and delayed-exponential models are very similar -- with the difference being mostly at very early epochs. Note that the delayed-exponential model also includes a burst, able to decouple old from recent SFH, but the optimal model has negligible burst fraction. This too is in agreement with the non-parametric model, which finds a very similar solution when adding more time bins at recent look-back times.
This spectacular agreement cannot be general \citep[see again e.g.,][]{leja+2019a}, and requires an explanation.
The most likely scenario is that the combination of weak Balmer break and moderate rest-UV flux requires a fairly constant SFR in the last few hundred Myr, which can be conveniently captured by both a delayed-exponential (with long e-folding time) and a non-parametric SFH with continuity prior.

Finally, we turn our attention to the AGN. Outside the high radio luminosity, we find no additional evidence of an AGN.
\hst imaging does not show a bright point source (Fig.~\ref{f.img.a}; \jwst imaging is severely affected by edge effects).
The presence of a Balmer break argues against a dominant AGN continuum at UV--optical wavelengths.

Similarly, we find no evidence for an upturn of the continuum at red wavelengths, around 8~\mum \citep[rest-frame 1.5~\mum; these wavelengths can already show evidence of AGN dust emission; e.g.,][although, admittedly, the sensitivity of our \textit{Spitzer}/IRAC observations is low at 6--8~\mum]{deugenio+2023c}.
Repeating the fiducial fit including an AGN dust torus in the model returns an AGN MIR luminosity consistent with 0.
All together, these lines of evidence argue against the presence of a radiatively efficient AGN, in agreement with most local radio galaxies \citep[e.g.,][]{heckman+best2014}, and as expected from the low-ionisation nebular emission.
However, in the local Universe, radiatively inefficient AGN are found mostly in high-mass radio-loud elliptical galaxies, while \target is clearly a gas-rich, lower-mass system.
This combination of LERG and star-forming host however agrees with observations, which find the fraction of quiescent-host radio-AGN decreases with redshift \citep{magliocchetti+2016} and at $z \gtrsim 1$ LERG hosts are predominantly star-forming galaxies, in agreement with the evolving fraction of quiescent galaxies  \citep{kondapally+2022}.

\subsection{A strong shock}\label{s.d.shock}

\target is known to harbour an AGN, due to its radio luminosity exceeding the limit of star formation \citep{pracy+2016}.
The emission-line ratios can only be explained by a strong shock.
The G395H spectrum, alone, is insufficient to adequately measure the properties of the host galaxy vs shock component. While we do find evidence of two components (with velocity dispersions 200 and~600~\kms; \S~\ref{s.obs.ss.highres}), these have unrealistic line ratios for the narrow component, therefore we cannot trust the resulting kinematics either.
Our joint modelling of the star-forming and shocked gas using both prism and G395H data prefers a fast shock solution ($v_\mathrm{sh} = 750$~\kms; Table~\ref{t.prosp} and Fig.~\ref{f.sed.a}). Assuming the typical conversion between de-projected velocity and line-of-sight velocity and velocity dispersion that is used in outflows \citep[i.e., $v_\mathrm{sh} = \lvert v_\mathrm{los}\rvert + 2\cdot\sigma_\mathrm{los}$; e.g.,][]{rupke+2005}, our $v_\mathrm{sh}$ is in loose agreement with the relatively high velocity dispersion of the gas, which we constrain independently from the G395H spectrum ($\sigma = 400\pm35$~\kms; Table~\ref{t.emlines} and Fig.~\ref{f.highres}).

As an alternative explanation, we ask whether virial equilibrium can explain the observed line broadening. To address this question, we compare the velocity dispersion from the virial theorem to the measured velocity dispersion \citep[we use the calibration of ][]{cappellari+2013b}.
With the stellar mass estimate of $10^{10.2}$~\Msun and a half-light radius of 2~kpc \citep[from the mass--size relation of][]{ward+2024}, we estimate a virial dispersion $\sigma_\mathrm{vir} = 90$~\kms -- clearly inconsistent with the observations. Overall, we can conclude that both the emission-line ratios and the gas kinematics argue for the presence of a strong shock.

The weakness of \OIIIL relative to lower-ionisation oxygen species tells us that this shock has run out of precursor gas, either by running through the entire galaxy ISM \citep[as seen in some local radio galaxies; e.g.,][]{drevet-mulard+2023}, or by `breaking free', i.e., crossing from the ISM to the low-density CGM.
The first interpretation seems more problematic, \emph{vis-\'a-vis} the model preference for solar metallicity; in this case, in fact, we would have a highly enriched galaxy at $z=4.6$ and at relatively low \citep[sub-knee, e.g.][]{weaver+2023} stellar mass, possibly in tension with scaling relations, as we have seen.
The other possibility -- the shock breaking free from the galaxy ISM -- seems more plausible; in this case, the shock metallicity would be tracing a relatively small central region of the galaxy, likely close to the centre.
Fast enrichment around AGN has been suggested in the past, and there are many processes which can expedite a fast-track chemical evolution for these regions \citep{baldwin+2003,cameron+2023a,huang+2023}.

\subsection{AGN feedback}\label{s.d.pjet}

As for the force driving the shock, the presence of a radio-loud AGN, coupled with the high shock velocity, strongly suggests an AGN origin. The clumpy nature of the rest-UV image (Fig.~\ref{f.img.a}) may suggest an ongoing merger, but UV morphology is often complex and is sensitive to the spatial distribution of even little foreground material, which masquerades the mass distribution.
Rest-optical \textit{Spitzer} imaging has insufficient spatial resolution to investigate the galaxy morphology, but NIRCam imaging from CEERS tentatively suggests a smooth rest-optical light distribution (Fig.~\ref{f.nircam}), so a major merger is disfavoured. A strong conclusion would require better image quality.

The lack of evidence for a strong AGN continuum (from X-rays, optical and MIR), coupled with the low-power radio luminosity ($P_{1.4\,\mathrm{GHz}} = 6.5\pm1.3 \times 10^{24}$W~Hz$^{-1}$), suggests a stage of radiatively inefficient accretion onto the supermassive black hole.
In the local Universe, this accretion mode has been interpreted as due to inefficient Bondi-Hoyle accretion of hot gas, supported by morphological evidence and low gas fractions of galaxies hosting low-power radio AGN \citep{hardcastle+2006,smolcic+riechers2011,heckman+best2014}.
\target certainly struggles to fit this hot-gas picture; its high SFR and low stellar mass imply a high gas fraction.

In terms of energy, it is difficult to accurately measure the total energy of the ionised gas, because -- under the assumption of an old shock -- an inordinate amount of gas could have already recombined, thus contributing little power to nebular emission.
Using the standard approach, we can estimate the mass and kinetic energy of the ionised gas \citep[e.g.,][]{venturi+2021} from the \Halpha luminosity, from the gas density and from the gas velocity dispersion. We infer the \Halpha luminosity from the \Halpha flux of the shock component in the model (Table~\ref{t.emlines.prosp}), and assume a density of $1,000$~cm$^{-3}$, giving a mass of $M_\mathrm{ion} \sim 2\times 10^7$~\Msun. We further assume a velocity dispersion of 400--600~\kms, spanning the range between the single-component fit and the broad component of the multiple-component fit (\S~\ref{s.obs.ss.highres}), which gives a kinetic energy of $E_\mathrm{kin} \sim 3\text{--}8 \times 10^{55}$~erg.
If we assume that the radio emission is due to a relativistic jet, we estimate the kinetic power of the jet to be $P_\mathrm{jet} = 4 \times 10^{44}$~erg~s$^{-1}$, by employing the local scaling relation between radio power and cavity (jet) power observed in cluster galaxies showing cavities in the X-ray haloes filled by radio emission \citep{cavagnolo+2010}.
This estimate comes with large uncertainties, though we note that the calibrations based on synchrotron emission and X-ray cavities tend to agree \citep{cavagnolo+2010,willner+2012,tadhunter2016}. Due to the low spatial resolution of the radio observations \citetext{3.8~arcsec; Fig.~\ref{f.img.a}; \citealp{ivison+2007}}, we are unable to study the spatial structure of the warm ionised gas relative to the radio emission, unlike in other high-redshift examples \citep{nesvadba+2017,saxena+2024}.
At face value, the above estimates of the gas mass and kinetic energy and jet power align with estimates from local, well resolved galaxies \citep{venturi+2021}.
This agreement suggests that the radio AGN is capable of powering the observed shock emission, though the uncertainties on our estimates are large. To estimate the total energy of the jet, we assume an age of $t_\mathrm{jet} < 6.5$~Myr, obtained by dividing the LOFAR FWHM by a speed of 0.01~c, corresponding to the speed of the advancing radio hotspot \citep{odea+2009}. This is again an order-of-magnitude estimate, due to the radio emission being spatially unresolved.
We lean toward an old shock age from the lack of nebular emission from high-ionisation species, which is characteristic of shocks without precursor gas. With this estimate, the ratio between the gas kinetic energy and the jet energy is $E_\mathrm{kin}/(P_\mathrm{jet}\cdot t_\mathrm{jet}) \sim 0.001$. 

The large energy difference between the AGN and shock implies a low-efficiency coupling between the AGN and the galaxy ISM -- as seen in most local radio-loud AGN.
Based on numerical simulations and observations \citep{mukherjee+2018,venturi+2021}, this inefficient coupling would require a near-polar jet alignment (relative to the galaxy disc).
The inefficient coupling agrees with the galaxy SFH, which we see increasing in the last 100~Myr, suggesting that AGN feedback did not alter the evolutionary trajectory of the galaxy. However, the large amount of jet energy freed from the black hole may couple with the galaxy CGM on halo-wide scales, seeding a quiescent future.

Alternatively, the SFR we infer may be connected to the AGN by positive feedback \citep[in agreement with many other cases, e.g.,][]{maiolino+2017,gallagher+2019}, for instance if the shock compressed the ISM and triggered increased star formation.
This scenario however seems unlikely, given that highly star-forming, dust-obscured emission is seen beyond the reach of the shock-driven emission (as traced by \OIIall, Fig.~\ref{f.img.b}).

In any case, radio-AGN feedback seems at best unable to interrupt ongoing star formation, and at worst, even promoting it. This argues against a major role of ejective feedback in \target.
On the other hand, radio AGN were likely already contributing to halo heating one Gyr after the Big Bang, starting the work needed to achieve preventive feedback.

Larger samples and more complete spatial coverage of this and similar sources may help clarify the role of radio AGN in the formation and evolution of galaxies in the first two billion years of the Universe.

\section{Summary and Conclusions}\label{s.conc}

In this work, we study the low-luminosity radio-loud AGN \target at redshift $z=4.6$, observed with \jwst/NIRSpec as part of the large GTO programme WIDE \citep{maseda+2024}.

The prism spectrum reveals a rich set of emission lines, including several metals; oxygen, nitrogen, sulphur and possibly carbon (Fig.~\ref{f.img.a}; Table~\ref{t.emlines}). Emission-line ratio diagnostics show high values for low-ionisation and neutral species, indicating a large reservoir of warm yet neutral gas, typical of shock-driven emission (Fig.~\ref{f.bpt}). The low luminosity of higher ionisation species (most notably \OIIIall) favours an old shock.

We study simultaneously the star-formation history and shock properties of \target under a Bayesian framework; we combine the grid of pre-computed shock models from the \href{https://sites.google.com/site/mexicanmillionmodels/}{database 3MdB} \citep{alarie+morisset2019} with the flexible galaxy inference tool \prospector \citep{johnson+2021}, using a static linear interpolator.
With this model, we are able to measure the properties of the AGN host (stellar mass and star-formation history) while taking into account any degeneracy with the shock parameters. The strongest degeneracies are between the shock luminosity fraction and the dust attenuation of the star-forming gas (Fig.~\ref{f.sed}).

We infer a low stellar mass $\log \mstar/\Msun = 10.16$~dex and an increasing SFH (Fig.~\ref{f.sed}; Table~\ref{t.prosp}), the latter suggesting that AGN feedback did not reduce the host star-formation rate.
These parameters are robust against different modelling assumptions -- even neglecting shocks -- supporting the conclusion that \mstar and SFR of low-luminosity radio-AGN hosts can be inferred successfully from SED modelling \citep{kondapally+2022}.
We find solar gas metallicity, which we tentatively interpret as evidence for fast-track chemical enrichment in the central regions of the host galaxy, probably related to the AGN.

The high shock velocity and the presence of a radio AGN suggest that the shock is AGN driven, similar to local radio-galaxies with radio jets.
The increasing SFH suggests that outflows, if present, have low mass loading and are unable to quench the galaxy, contrary to what observed in quenching galaxies at $z=2\text{--}3$ \citep[e.g.,][]{belli+2023,deugenio+2023c,davies+2024}.
The low kinetic energy of the ionised gas (relative to the radio power), as well as the increasing SFH, both suggest the coupling between the radio-AGN and the galaxy ISM to be inefficient.
This would leave a large amount of energy to be ultimately dumped into the galaxy halo, a necessary step towards achieving preventive feedback, which is suggested by both observations \citep{terrazas+2017,bluck+2022,bluck+2024} and by theoretical models \citep[e.g.,][]{piotrowska+2022}.

Our observations showcase the amazing capability of WIDE to reveal the detailed properties of bright, rare sources.
Deeper observations may reveal additional emission lines, enabling a more detailed characterisation of the physical properties of this system.
Spatially resolved spectroscopy with NIRSpec/IFS would greatly improve our understanding of the geometry of this system.

%-------------------------------------------------------------------

%%%%%%%%%%%%%%%%%%%%%%%%%%%%%%%%%%%%%%%%%%%%%%%%%%

%%%%%%%%%%%%%%%%% BODY OF PAPER %%%%%%%%%%%%%%%%%%

\section*{Acknowledgements}
The authors are grateful to Bernd Husemann for helping with the initial target selection of WIDE -- particularly for AGN.
FDE, RM, XJ and JS acknowledge support by the Science and Technology Facilities Council (STFC), by the ERC through Advanced Grant 695671 ``QUENCH'', and by the
UKRI Frontier Research grant RISEandFALL. RM also acknowledges funding from a research professorship from the Royal Society.
GM acknowledges financial support from the grant PRIN MIUR 2017PH3WAT (`Black hole winds and the baryon life cycle of galaxies').
SC and GV acknowledge support by European Union's HE ERC Starting Grant No. 101040227 - WINGS.
MVM is supported by the National Science Foundation via AAG grant 2205519, the Wisconsin Alumni Research Foundation via grant MSN251397, and NASA via STScI grant JWST-GO-4426.
AJB and GCJ acknowledge funding from the ``FirstGalaxies'' Advanced Grant from the European Research Council (ERC) under the European Union's Horizon 2020 research and innovation program (Grant agreement No. 789056).
ST acknowledges support by the Royal Society Research Grant G125142.
CT acknowledges support from STFC grants ST/R000964/1 and ST/V000853/1.
H{\"U} gratefully acknowledges support by the Isaac Newton Trust and by the Kavli Foundation through a Newton-Kavli Junior Fellowship.
This study makes use of data from AEGIS, a multiwavelength sky survey conducted with the Chandra, GALEX, Hubble, Keck, CFHT, MMT, Subaru, Palomar, Spitzer, VLA, and other telescopes and supported in part by the NSF, NASA, and the STFC.
The imaging data products presented herein were retrieved from the Dawn JWST Archive (DJA). DJA is an initiative of the Cosmic Dawn Center (DAWN), which is funded by the Danish National Research Foundation under grant DNRF140.

LOFAR is the Low Frequency Array designed and constructed by ASTRON. It has observing, data processing, and data storage facilities in several countries, which are owned by various parties (each with their own funding sources), and which are collectively operated by the ILT foundation under a joint scientific policy. The ILT resources have benefited from the following recent major funding sources: CNRS-INSU, Observatoire de Paris and Universit\'e d'Orl\'eans, France; BMBF, MIWF-NRW, MPG, Germany; Science Foundation Ireland (SFI), Department of Business, Enterprise and Innovation (DBEI), Ireland; NWO, The Netherlands; The Science and Technology Facilities Council, UK; Ministry of Science and Higher Education, Poland; The Istituto Nazionale di Astrofisica (INAF), Italy.

This research made use of the Dutch national e-infrastructure with support of the SURF Cooperative (e-infra 180169) and the LOFAR e-infra group. The J\"ulich LOFAR Long Term Archive and the German LOFAR network are both coordinated and operated by the J\"ulich Supercomputing Centre (JSC), and computing resources on the supercomputer JUWELS at JSC were provided by the Gauss Centre for Supercomputing e.V. (grant CHTB00) through the John von Neumann Institute for Computing (NIC).

This research made use of the University of Hertfordshire high-performance computing facility and the LOFAR-UK computing facility located at the University of Hertfordshire and supported by STFC [ST/P000096/1], and of the Italian LOFAR IT computing infrastructure supported and operated by INAF, and by the Physics Department of Turin university (under an agreement with Consorzio Interuniversitario per la Fisica Spaziale) at the C3S Supercomputing Centre, Italy.

This work made extensive use of the freely available \href{http://www.debian.org}{Debian GNU/Linux} operative system.
We used the \href{http://www.python.org}{Python} programming language \citep{vanrossum1995}, maintained and distributed by the Python Software Foundation. We made direct use of Python packages
{\sc \href{https://pypi.org/project/astropy/}{astropy}} \citep{astropyco+2013},
{\sc \href{https://pypi.org/project/corner/}{corner}} \citep{foreman-mackey2016},
{\sc \href{https://pypi.org/project/dynesty/}{dynesty}} \citep{speagle2020,koposov+2023},
{\sc \href{https://pypi.org/project/emcee/}{emcee}} \citep{foreman-mackey+2013},
{\sc \href{https://pypi.org/project/jwst/}{jwst}} \citep{alvesdeoliveira+2018},
{\sc \href{https://pypi.org/project/matplotlib/}{matplotlib}} \citep{hunter2007},
{\sc \href{https://pypi.org/project/numpy/}{numpy}} \citep{harris+2020},
{\sc \href{https://pypi.org/project/pingouin/}{pingouin}} \citep{vallat2018},
{\sc \href{https://pypi.org/project/ppxf/}{ppxf}} \citep{cappellari+emsellem2004, cappellari2017, cappellari2022},
{\sc \href{https://pypi.org/project/astro-prospector/}{prospector}} \citep{johnson+2021} \href{https://github.com/bd-j/prospector}{v2.0},
{\sc \href{https://pypi.org/project/PyNeb/}{pyneb}} \citep{luridiana+2015},
{\sc \href{https://pypi.org/project/python-fsps/}{python-fsps}} \citep{johnson_pyfsps_2023},
{\sc \href{https://pypi.org/project/scipy/}{scipy}} \citep{jones+2001}, and {\sc \href{https://github.com/AstroJacobLi/smplotlib}{smplotlib}} \citep{smplotlib}.
{\sc dynesty} uses the nested sampling algorithms of \citet{skilling2004,skilling2006} and \citet{higson+2019}.
\shinbnzv uses the {\sc \href{http://www.qhull.org/}{qhull}} triangulation algorithm \citep{barber+1996}.
We also used the softwares {\sc \href{https://github.com/cconroy20/fsps}{fsps}} \citep{conroy+2009,conroy_gunn_2010}, {\sc \href{https://www.star.bris.ac.uk/~mbt/topcat/}{topcat}}, \citep{taylor2005}, {\sc \href{https://github.com/ryanhausen/fitsmap}{fitsmap}} \citep[known as `mapviewer' in the DJA;][]{hausen+robertson2022} and {\sc \href{https://sites.google.com/cfa.harvard.edu/saoimageds9}{ds9}} \citep{joye+mandel2003}.
The data reduction on the DJA makes use of {\sc \href{https://pypi.org/project/grizli/}{grizli}}
\footnote{\href{10.5281/zenodo.1146904}{10.5281/zenodo.1146904}}.

%%%%%%%%%%%%%%%%%%%%%%%%%%%%%%%%%%%%%%%%%%%%%%%%%%
\section*{Data Availability}

The reduced and calibrated spectra are available as part of the first public data release of WIDE \citep{maseda+2024}.
Reduced images presented herein were retrieved from the Dawn JWST Archive (DJA). DJA is an initiative of the Cosmic Dawn Center (DAWN), which is funded by the Danish National Research Foundation under grant DNRF140.
The photometric measurements used in the \prospector fits are available from \citet{stefanon+2017}.

This work is based on observations made with the NASA/ESA/CSA James Webb Space Telescope. The data were obtained from the \href{https://mast.stsci.edu/portal/Mashup/Clients/Mast/Portal.html}{Mikulski Archive for Space Telescopes} at the Space Telescope Science Institute, which is operated by the Association of Universities for Research in Astronomy, Inc., under NASA contract NAS 5-03127 for JWST. These observations are associated with programmes PID~1213 (NIRSpec WIDE MOS Survey -- AEGIS; PI~N.~L\"utzgendorf)
and PID~1345 (The Cosmic Evolution Early Release Science (CEERS) Survey; PI~S.~L.~Finkelstein).

The software developed for this work will be made available after review. It includes the \prospector derived classes \texttt{ShockSpecModel} and \texttt{ShockPolySpecModel}, and a static implementation of \shinbnzv.
At the time of this writing, the documentation is wanting, but we encourage the reader to address questions to the \sendemail{fdeugenio@gmail.com}{Questions about shinbnzv}{Dear Francesco,\%0A\%0Ahow are you? I have a question about the code "shinbnzv" (sic), if I may.\%0AThe thing is, that ...\%0A\%0ARegards,\%0A}{package maintainers}.
The package shows how to build a custom interpolator model to use different input grids and different emission-line selections; this requires access the library of pre-computed shock models from the \href{https://sites.google.com/site/mexicanmillionmodels/}{database 3MdB} \citep{alarie+morisset2019}, which can be requested to the database maintainers.
 


\bibliographystyle{config/mnras}
\bibliography{shockz5.bbl} % if your bibtex file is called example.bib

\appendix

\section{SED modelling without shocks}\label{a.sfonly}

As a reference, we measure the physical properties of \target without modelling the shock-driven nebular emission.
We model simultaneously the photometry and prism spectrum using the standard version of \prospector \citep{johnson+2021}.
As we have seen (\S~\ref{s.sh}--\ref{s.host}), half of the nebular emission in \target is due to shocks.

\begin{figure}
  \includegraphics[width=\columnwidth]{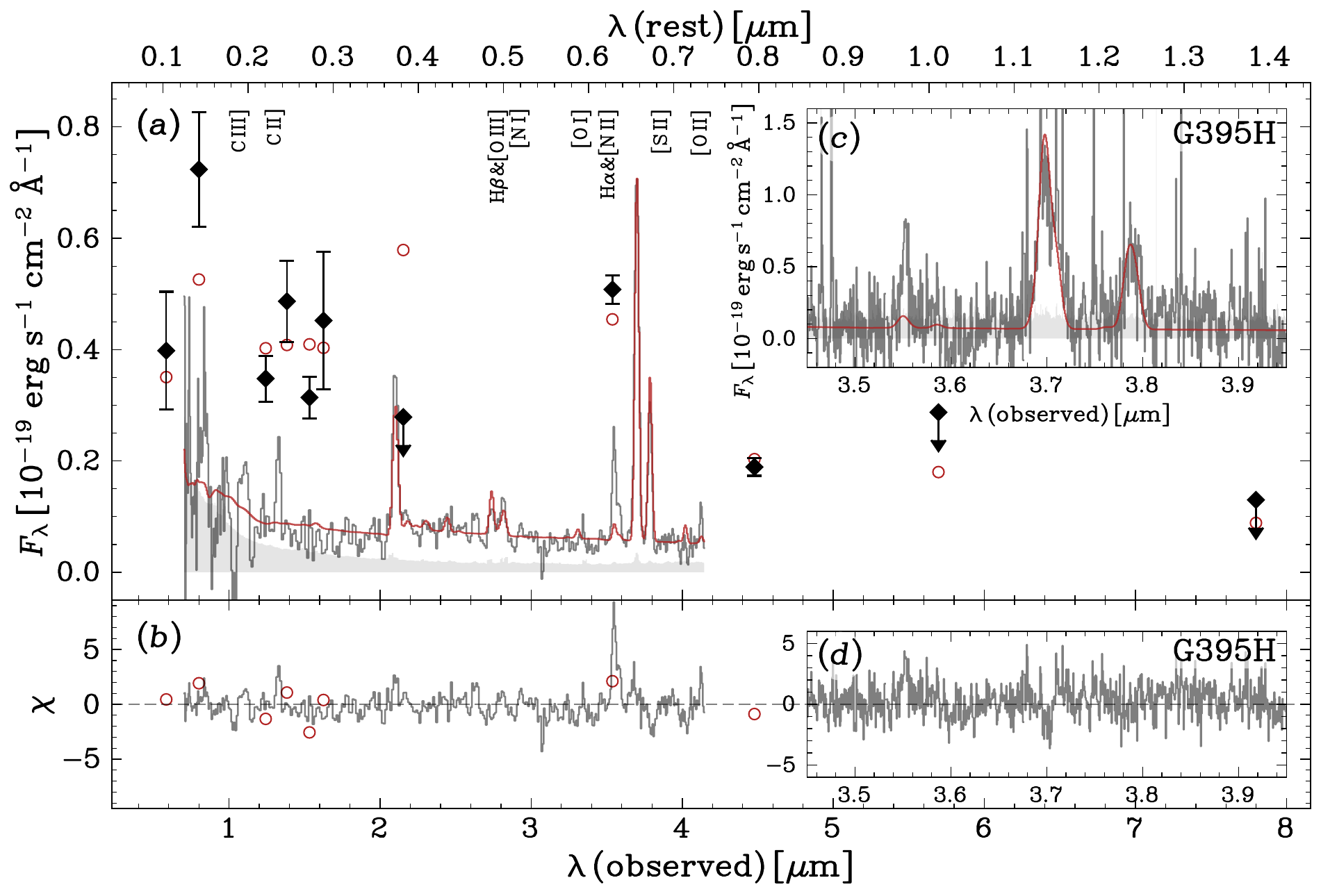}
  {\phantomsubcaption\label{f.a.prosp.a}
   \phantomsubcaption\label{f.a.prosp.b}
   \phantomsubcaption\label{f.a.prosp.c}}
  \caption{Modelling the data without including shocks we find very similar \mstar as the fiducial case, but this is by happenstance; the model clearly mismatches some of the observed emission-line ratios, particularly \OIIall, and \Halpha+\NIIall,
  resulting in an overestimated stellar continuum. The mismatch is partly due to the competing requirements to have high dust attenuation (to reproduce the observed Balmer decrement) and high \OIIall flux, which requires low attenuation, for reasonable metallicity. \OIall is also not reproduced; its observed ratio with \Halpha is beyond the capabilities of star-forming models.
  The lines and symbols are the same as Fig.~\ref{f.sed.b}--\subref{f.sed.e}.
  }\label{f.a.prosp}
\end{figure}

We use the \prospector \texttt{PolySpecModel}, with the same model parameters and prior probability distributions of Table~\ref{t.prosp}, except for the shock-specific parameters, which are absent.
The resulting model is shown in Fig.~\ref{f.a.prosp}; as expected, the model fails to reproduce the flux of high-temperature and/or low-ionisation emission, namely \CIIall, \OIIall, \OIall and \OIIAuall.
The observed line strengths are so far from what a star-forming model can generate, that the optimisation resorts to increasing the level of the continuum to reduce the residuals (Fig.~\ref{f.a.prosp.b}, around wavelengths 1.5 and 3.2--3.3~\mum).
In addition, the model cannot reproduce the observed Balmer decrement, tending to both underestimate \Halpha and overestimate \Hbeta. This is due to the fact that higher dust attenuation would make it impossible to reproduce the already under-estimated flux of \OIIall.

Despite this shortcoming, the resulting model has $\log\,\mstar [\Msun] = 10.1\pm0.1$~dex and $\log\,\mathrm{SFR}_{100} [\Msun~\peryr] = 1.6\pm0.1$~dex, statistically consistent with the fiducial value.
The agreement in SFR is remarkable, given that in this model the emission lines are entirely due to star formation, whereas in the fiducial model 50~per cent of the \Hbeta luminosity is due to shocks (Table~\ref{t.prosp}). However, for the star-forming model, the higher \Halpha and \Hbeta luminosities are compensated by lower dust attenuation than in the fiducial model with shocks ($\tau_V = 0.58\pm0.06$ and $\mu=0.5\pm0.2$), hence the total SFR is still very close to the fiducial value.
Similarly, the agreement in \mstar is due to the combination of lower dust attenuation (which decreases \mstar) and a systematically over-predicted continuum level (which increases \mstar).

Compared to previous studies, we find a lower \mstar; we trace this difference to the inclusion of the spectrum in our analysis.
Indeed, using photometry only to constrain the above model, we too find a higher value, $\log \mstar [\Msun] = 10.4\pm0.1$, the same as the literature value of 10.4 \citep{stefanon+2017}.

\section{Shock interpolator}\label{a.interp}

The interpolator we have built for this study predicts the flux of various emission lines.
Specifically, we retrieve the line fluxes from the 3MdB database, then construct a grid in \textit{BnZv} space. We transform $B$, $n$ and $Z$ from linear to log-space (\S~\ref{s.sh.ss.shint}).
To interpolate, we use the {\sc qhull} algorithm \citep{barber+1996} to triangulate the input grid, then perform linear interpolation between the barycentre of each triangle.

The specific interpolator realisation we used in this work predicts the logarithm of the flux of 52 emission lines.
These include the brightest recombination lines in the rest UV--NIR range (i.e., hydrogen and helium), strong, non-resonant metal lines, and the brightest metal auroral lines.
Notable lines that we do not model are \Lyalpha, \CIVall and \MgIIall. Aside from these three resonant transitions, all strong lines from \HeIIL[1640] to \Paalpha are modelled.
In addition, we model the low-ionisation \NIall and the temperature-sensitive auroral lines \CIIall, \SIIL[4070], \OIIIL[4363], \OIL[5577], \NIIL[5755], and \OIIAuall.
The interpolator itself does not apply any dust reddening, nor does it attempt to model the spectral profile of the lines; where relevant in this work, dust reddening and line broadening are delegated to the caller program.

This specific interpolator can be easily generalised to use more or less lines, limited only by their availability in 3MdB.

\begin{figure}
  \includegraphics[width=\columnwidth]{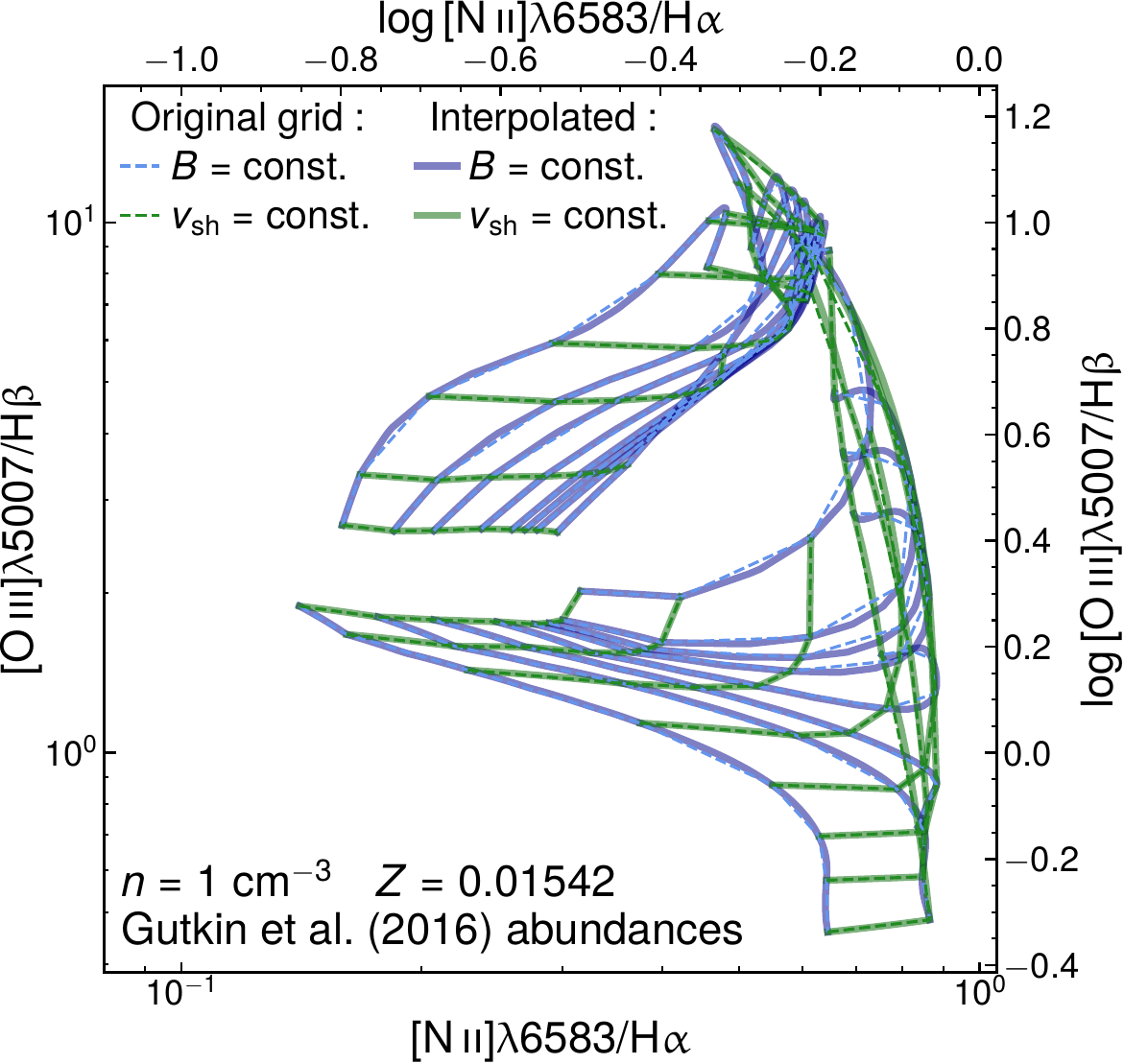}
  \caption{Our linear interpolator \shinbnzv correctly predicts the BPT line ratios from the input grid, down to the machine precision. We show the grid
  \texttt{Gutkin16} from the \href{https://sites.google.com/site/mexicanmillionmodels/}{database 3MdB} \citep{alarie+morisset2019}.
  We selected the model with $Z=0.01542$ and pre-shock density $n_\mathrm{sh}=1$~cm$^{-3}$, with abundances from \citet{gutkin+2016}.
  The thin dashed lines are the original grid from 3MdB,
  the thick solid lines are the grid interpolated with \shinbnzv.
  Mismatches between the two grids are due to the choice of interpolating
  linearly, and are caused by the different rate of change of the numerator and denominator.
  The grids show eight logarithmically spaced values of magnetic field strength $B$
  and shock velocity $v_\mathrm{sh}$.
  The top and bottom grids show respectively the model including a precursor and the shock-only
  model.}\label{a.shinbnzv}
\end{figure}

To evaluate the accuracy of the interpolator, we compare the input grids from 3MdB (dashed lines) to the grids predicted by \shinbnzv (solid lines), on the BPT diagram (Fig.~\ref{a.shinbnzv}).
By construction, the model predicts the emission-line fluxes exactly at every grid node; deviations between the input and interpolated ratios are due to the common practice of connecting linearly between grid nodes in the 2-d space of emission-line ratios.

\section{Pure shock model}\label{a.shockonly}

A shock-only model -- without photoionisation due to star formation -- cannot explain the line ratios observed in \target.
Here we consider the shock models of \S~\ref{s.sh.ss.shint}, and fit eight emission-line ratios, taking the measurements from Table~\ref{t.emlines}: \OIIall/\Hbeta, \OIIIall/\Hbeta, \OIall/\Hbeta, \NIIL/\Halpha, \Halpha/\Hbeta, \SIIall/\Halpha, \SIIL[6716]/\SIIL[6731] and \OIIAuall/\OIIall.
We tested shock plus precursor models, but found that shock-only models agreed much better with the data, as discussed in \S~\ref{s.sh.ss.bpt}.
For this reason, here we report just the shock-only models.
Our setup consists in combining the emission-line ratios from the shock models with the \citet{cardelli+1989} attenuation curve, parametrised by the $V$-band attenuation $A_V$.
This model has five free parameters: $A_V$ and the four shock parameters already introduced in \S~\ref{s.sh.ss.shint} (like for the \prospector model, we rescale the metallicity values to $\Zsun=0.0142$).
We use a Bayesian approach, where the probability of observing each ratio is Gaussian.
The prior probability distribution of the shock parameters are flat in log space, spanning the range of parameters from the grid.
The prior of $A_V$ is flat between $0\leq A_V \leq 4$~mag.
The posterior probability is integrated using \emcee \citep{foreman-mackey2016}, and at each step the likelihood is calculated using the \shinbnzv interpolator.

\begin{figure}
\includegraphics[width=\columnwidth]{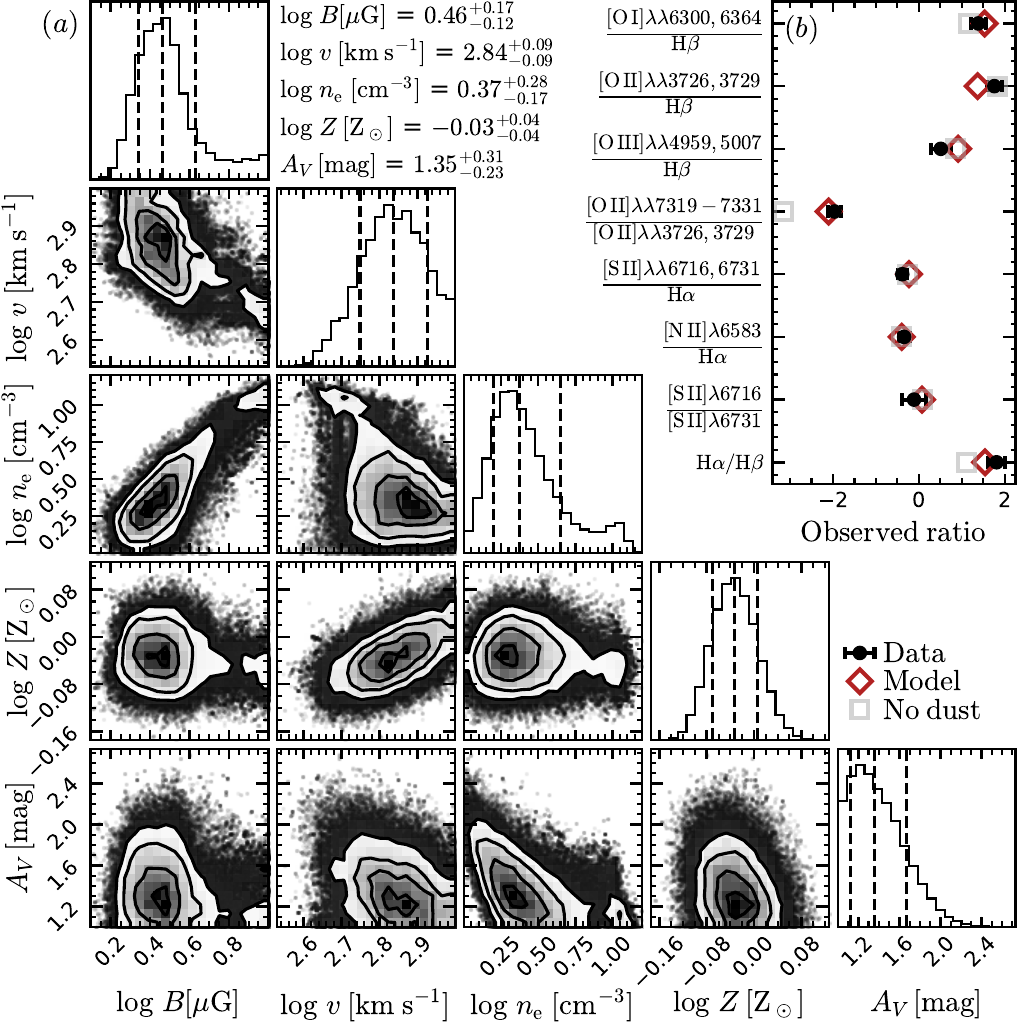}
{\phantomsubcaption\label{f.c.shock.a}
 \phantomsubcaption\label{f.c.shock.b}}
\caption{Marginalised posterior probability distribution for the shock-only model to the observed emission-line ratios. Panel~\subref{f.c.shock.b} shows the observed line ratios (black), predicted ratios (red diamonds) and de-reddened ratios (grey squares).
Without star formation, shocks are unable to simultaneously reproduce all the observed ratios (reduced $\chi^2=3.9$), due to the incompatible requirements of high observed \OIIall/\Hbeta and \Halpha/\Hbeta.
Overall, the best-fit shock parameters are very similar to the results of the \prospector composite model, which combines shocks and star formation.}\label{f.c.shock}
\end{figure}

The results are shown in Fig.~\ref{f.c.shock}; the corner diagram displays the marginalised posterior distribution, exhibiting the known degeneracy between the magnetic field strength $B$ and the pre-shock density $n$.
Panel~\subref{f.c.shock.b} shows the data (black circles) and the model prediction (red diamonds); we also report the intrinsic, de-reddened line ratios as grey squares.
The posterior parameter values are all reasonably close to the \prospector model from \S~\ref{s.host.ss.prosp}.
But despite this agreement, it is clear that the shock model cannot satisfactorily reproduce both the low-ionisation species and the strong Balmer decrement; the outcome is an under-predicted \Halpha/\Hbeta ratio.
Note that this solution is driven primarily by the strong prior in $A_V$; allowing $A_V$ to assume values higher than $2$~mag, the model favours high dust attenuation (thus matching the Balmer decrement) but at the expense of the \OIIall/\Hbeta ratio, which is then under-predicted.
Between these alternatives, it is clear that the low-attenuation solution is the most credible, because the fraction of the \Halpha flux that is unexplained by the shock can easily be ascribed to dust-obscured star formation.
In contrast, the high-attenuation solution leaves unexplained the \OIIall flux, which is hard or even impossible to produce without also altering the other line ratios.

A final word of caution is that -- in the presence of different physical mechanisms powering nebular emission -- the \Halpha/\Hbeta ratio may not necessarily give an unbiased estimate of dust attenuation.

\section{Elementary recovery tests}\label{a.recovery}

\begin{figure}
  \includegraphics[width=\columnwidth]{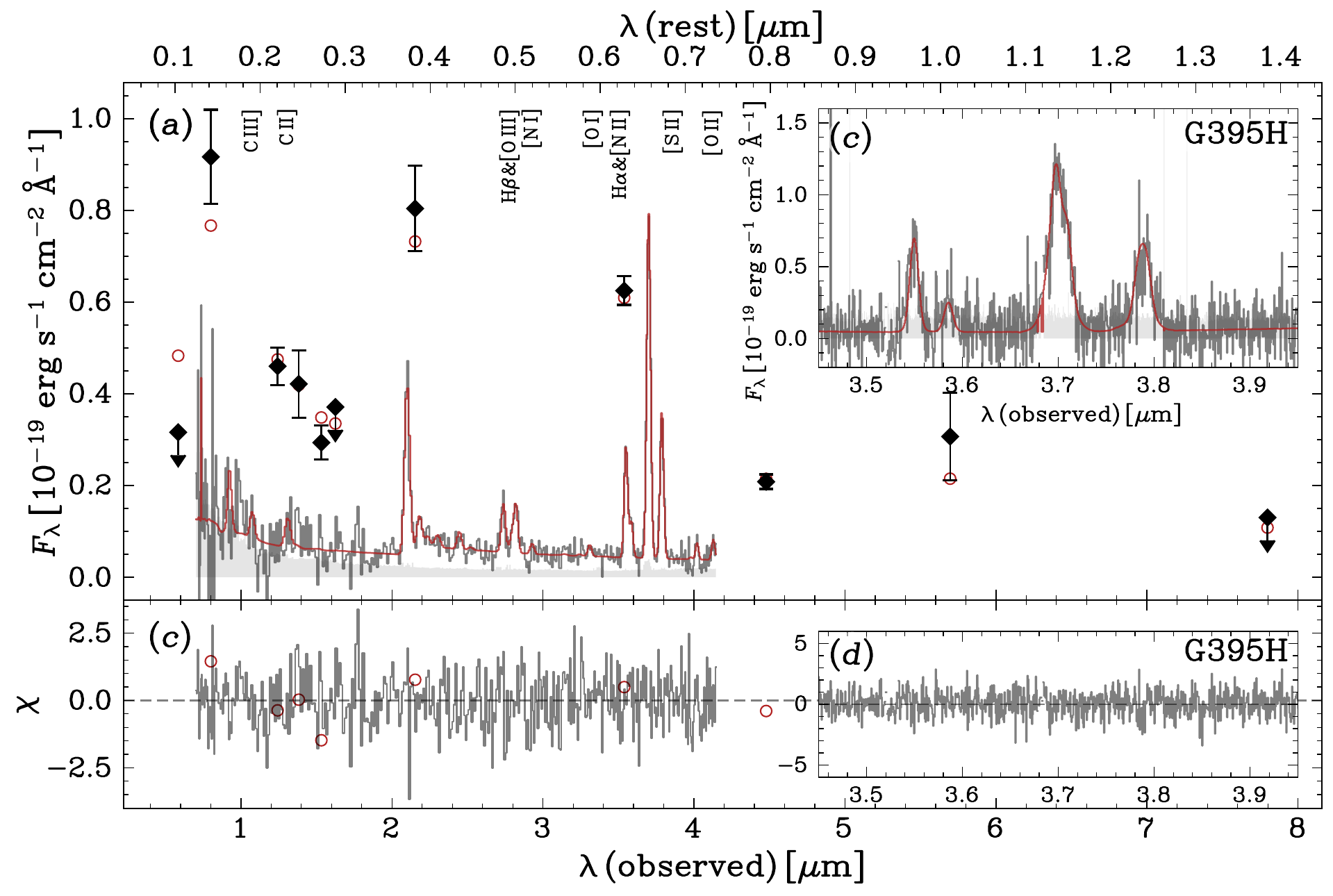}
  \caption{Mock observation and re-fit of the fiducial model (Table~\ref{t.prosp} and Fig.~\ref{f.sed}), with input parameters equal to fiducial model posterior, but setting
  $\log\,L(\Hbeta)_\mathrm{sh}=0.06$. The symbols and panels are the same as Fig.~\ref{a.sfonly}. Overall, the mock data (grey) is of similar quality as the WIDE observations (cf.~Fig.~\ref{f.sed.b}--\subref{f.sed.d}).
  Notable differences are the absence of outliers in the G395H spectrum (outliers are masked in the analysis), and the much higher $K_s$-band flux.
  }\label{a.mockspec}
\end{figure}

To assess the ability of the software to infer the input parameters, we conduct a set of simple recovery tests.
Due to the computational cost of these tests, we focus on a single model, the fiducial \prospector model present in \S~\ref{s.host.ss.prosp}.
We create mock observations using the fiducial model parameters listed in Table~\ref{t.prosp}, and varying one parameter at a time, as follows:
$\log\,L(\Hbeta)_\mathrm{sh}=0.06$, $\log\,v_\mathrm{sh}=2.6$, 
$\log\,\mstar=11.3$, $\log\,Z_\mathrm{gas}=-0.5$, 
$\tau_V = 0.0$, $\tau_V=1.3$, $\log\,n_\mathrm{gas}=2$, and
$\log\,B_\mathrm{gas}=-1$ (all units are as in Table~\ref{t.prosp}).
The mocks are generated by first creating the model spectrum, then matching the spectral resolution and wavelength grid of the observations, and finally adding random noise to mimic the noise of the input data.
For photometry, we draw the mock data from a Gaussian distribution with mean equal to the model flux and standard deviation equal to the observed noise.
For the spectra, noise is generated by adding the residuals of the fiducial model to the mock spectrum; for each spectral pixel, we randomly draw the residuals in a narrow wavelength window around the pixel (but we do not use residuals identified as outliers).
An example mock model is shown in Fig.~\ref{a.mockspec}.

Overall, we find that the input parameters are recovered within 1~\textsigma, except for $\log\,B_\mathrm{sh}$ and $\log\,n_\mathrm{sh}$, which tend to have larger excursions of 1.5~\textsigma.
Physically, the strongest constraint on these two parameters comes from multi-level forbidden line sets. Unfortunately, none of these emission features are spectrally resolved in the prism spectrum. In the G395H spectrum, the only detection useful for disentangling
$\log\,B_\mathrm{sh}$ and $\log\,n_\mathrm{sh}$ is the \SIIall doublet, but the pernicious combination of high dispersion and low signal-to-noise ratio prevents us from drawing strong constraints; indeed, the posterior probability distribution of the \SIIL[6716]/\SIIL[6731] ratio in Fig.~\ref{f.highres} spans the full allowed range.

Overall, the outcome of our tests confirm that the software implementation is self consistent, and that the parameter recovery process does not introduce bias.

% Don't change these lines
\bsp	% typesetting comment
\label{lastpage}
\end{document}